\documentclass[journal]{IEEEtran}
\IEEEoverridecommandlockouts
\usepackage{cite}
\usepackage{amsmath,amssymb,amsfonts}
\usepackage{algorithmic}
\usepackage{textcomp}
\usepackage{xcolor}
\usepackage{graphicx}
\usepackage{epstopdf}
\usepackage{float}
\usepackage[tight,TABTOPCAP]{subfigure}
\usepackage{multirow}
\usepackage{diagbox} 
\usepackage{makecell}
\usepackage{amsmath}
\usepackage{array}
\usepackage{tabu}
\usepackage{url}
\usepackage{threeparttable}
\usepackage[T1]{fontenc}
\usepackage{xcolor,cite,etoolbox}
\usepackage{booktabs}
\makeatletter
\pretocmd\@bibitem{\color{black}\csname keycolor#1\endcsname}{}{\fail}
\newcommand\citecolor[1]{\@namedef{keycolor#1}{\color{red}}}
\makeatother
\newcolumntype{I}{!{\vrule width 3pt}}
\newlength\savedwidth

\newlength\savewidth

\def\BibTeX{{\rm B\kern-.05em{\sc i\kern-.025em b}\kern-.08em
    T\kern-.1667em\lower.7ex\hbox{E}\kern-.125emX}}

\begin{document}

\title{Raw Bayer Pattern Image Synthesis for Computer Vision-oriented Image Signal Processing Pipeline Design}


\author{Wei Zhou,~\IEEEmembership{Student Member,~IEEE}, Xiangyu Zhang, Hongyu Wang, Shenghua Gao and Xin Lou, ~\IEEEmembership{Member,~IEEE} 
\thanks{The authors are with the School of Information Science and Technology, ShanghaiTech University, Shanghai, 201210, China.}}

\maketitle

\begin{abstract}

In this paper, we propose a method to add constraints that are un-formulatable in generative adversarial networks (GAN)-based arbitrary size RAW Bayer image generation. It is shown theoretically that by using the transformed data in GAN training, it is able to improve the learning of the original data distribution, owing to the invariant of Jensen–Shannon (JS) divergence between two distributions under invertible and differentiable transformation. Benefiting from the proposed method, RAW Bayer pattern images can be generated by configuring the transformation as demosaicing. It is shown that by adding another transformation, the proposed method is able to synthesize high quality RAW Bayer images with arbitrary size. Experimental results show that images generated by the proposed method outperforms the existing methods in the Fréchet inception distance (FID) score, peak signal to noise ratio (PSNR) and mean structural similarity (MSSIM), and the training process is more stable. To the best knowledge of the authors, there is no open-source, large-scale image dataset in the RAW Bayer domain, which is crucial for research works aiming to explore the image signal processing (ISP) pipeline design for computer vision tasks. Converting the existing commonly used color image datasets to their corresponding RAW Bayer versions, the proposed method can be a promising solution to the RAW image dataset problem. We also show in the experiments that, by training object detection frameworks using the synthesized RAW Bayer images, they can be used in an end-to-end manner (from RAW images to vision tasks) with negligible performance degradation. 
\end{abstract}


\section{Introduction}
\label{section:sec1}
Deep learning based computer vision algorithms have achieved impressive performance in various fields. In most of the applications, image sensors provide almost all the information consumed by vision algorithms. Modern consumer cameras usually employ mono-sensor technique to generate three-channel color images, where a color filter array (CFA) such as Bayer pattern array\cite{RN4} is covered on the image sensor to restrict each pixel captures only one color component among the red, green and blue. A step-by-step chain of image processing operations, usually referred to as image signal processing (ISP) pipeline, is then applied to generate a finely-rendered color image.

As a matter of course, most of the modern computer vision algorithms, take color images generated by ISP pipelines as input. The suitability of processed color images (instead of RAW images from image sensors) for computer vision applications is seldom questioned, though it is well-known that ISP pipelines may undermine the original information captured by the image sensors \cite{RN14}. The most critical reason is that all the well labeled large-scale datasets are in the format of processed, three-channel color images. In many applications, it is  assumed that images are directly related to the actual scene radiance, with cameras being accurate light measuring devices. However, it is well-known that ISP pipelines used in digital cameras are usually non-linear. They are designed for photography with a goal of generating high-quality images for human consumption, rather than capturing accurate physical descriptions of the scene. Therefore, the output pixel intensity values are non-linearly related to spectral scene radiance for most of the consumer cameras\cite{RN61,RN62,RN63}. On the other hand, the experimental results in \cite{RN61} show that the output RAW values of image sensors are linearly related to image irradiance and many visual tasks can benefit from these linearity properties. However, compensating the nonlinearities of processed color images from consumer cameras is difficult, since ISP pipelines of different camera manufactures are confidential and will not release to users. Moreover, it is intuitive that the optimal ISP pipeline configuration, i.e., necessary stages and their orders, for a specific vision task, may be different. Therefore, instead of using a standard ISP pipeline, it is necessary to determine the configuration of the ISP pipeline according to the vision tasks. In other words, co-design of ISP pipeline and vision algorithm is the key for further improvement of computer vision performance.

To take advantage of the linearity of RAW images and explore ISP configuration for specific tasks, the commonly used stages in ISP pipelines have been studied for both traditional and deep learning based computer vision algorithms \cite{RN59,RN23,RN56,RN57,RN55,RN64}. Due to the high complexity of modern ISP pipelines, sufficiently large number of RAW Bayer image data is needed for the study of configuration of ISP pipelines for different computer vision tasks. However, almost all the large-scale datasets are provided in the format of processed color image and there is few RAW Bayer image datasets. Although there are options to save RAW Bayer images for both digital single lens reflex (DSLR) cameras and mobile phones\cite{RN65}, it takes a huge amount of efforts to produce a large-scale RAW Bayer image dataset like ImageNet \cite{RN75}. To the best knowledge of the authors, there are only some small-scale RAW Bayer image datasets available\cite{RN55,RN64}. But they are produced for traditional computer vision algorithms like support vector machine (SVM), such that the volume of these datasets are too small for deep learning-based algorithms. Therefore, it is crucial to generate a RAW Bayer image dataset to explore the co-design of ISP pipelines and vision algorithms.

Another interesting application of RAW Bayer image dataset is in smart image sensor, where the image sensor and back-end computing engine are integrated together in one chip. This paradigm is always referred to as in-sensor computing\cite{RN78}. For in-sensor machine vision applications, if the ISP pipeline is completely bypassed, the performance of the algorithms may degrade by as much as 60\% according to \cite{RN70}. This is because the models are trained on finely rendered color image datasets, while the input images for inference are RAW Bayer pattern images. However, if the models can be trained or fine-tuned on RAW Bayer image datasets, this kind of performance degradation may be avoided, or at least alleviated.

For RAW image dataset generation, apart from capturing and labeling directly, reversible pipeline is another solution to generate RAW Bayer images from their color versions\cite{RN61,RN62,RN63}. One example is the configurable \& reversible imaging pipeline (CRIP) tool proposed in \cite{RN23}, where an ISP pipeline is modeled as serious of reversible transformations, such that finely rendered color images can be converted back to RAW Bayer images by reverse transformations. Motivated by the reversible ISP pipelines as well as the success of generative adversarial networks (GAN) in image-to-image translation \cite{RN66,RN67,RN68,RN69}, we propose in this work to transform a finely rendered color image to its corresponding RAW Bayer image pixel by pixel directly, where the color-to-RAW transformation is modeled as an image-to-image translation problem. Since the RAW Bayer images have their specific CFA patterns, e.g., the Bayer pattern, GAN cannot be directly applied to color-to-RAW image transformation, otherwise the color information will be lost and undesirable effects will be produced. Therefore, we need to maintain the CFA patterns during the transformation to avoid loose of color information.

In this paper, we propose a color-to-RAW conditional-GAN (cGAN)  to convert finely rendered color images to their corresponding RAW Bayer images, where an improved loss function is used to maintain the CFA structure. The contributions of this work are as follows:
\begin{itemize}
    \item Propose a method to add un-formulatable constraints in image generation (e.g., to generate specific patterns), and implement the proposed method on an end-to-end GAN model to convert finely rendered color images to RAW Bayer images in a pixel to pixel manner.
    
    \item Introduce large-scale RAW Bayer datasets for works which aim to find the optimal ISP configurations for specific computer vision tasks. These datasets are reverted version of standard datasets that are commonly used in the computer vision field.
    \item Provide a new solution to the mismatch between testing set and training set for the in-sensor computing paradigm \cite{RN70}.
\end{itemize}


\section{Related Work and Applications}
\subsection{Radiometric Calibration \& Reversible Camera Model}
In the field of computer vision, radiometric calibration refers to the process of converting the irradiance received by an image sensor to pixel intensity. Conventional ``grayscale'' radiometric calibration\cite{RN71,RN72,RN73} can be expressed as
\begin{equation}
    \label{gray_scale_calibration}
    I\left ( x,y \right )=f\left ( \kappa \left ( x,y \right ) \right ).
\end{equation}
Here, $\kappa (x,y)$ is the camera's RAW output at pixel location $(x,y)$, which is linearly related to the image irradiance, and $I(x,y)$ is the corresponding finely rendered pixel intensity. The function $f(\cdot)$ is the radiometric response function, which is assumed to be a fixed property of a camera. To deal with color images, we can apply the model in \eqref{gray_scale_calibration} to each channel independently, such that $I$ can be extended as     
\begin{equation}
    \begin{split}
    \label{color_calibration1}
    I(x,y)&=\begin{bmatrix}
        I_r(x,y)\\
        I_g(x,y)\\
        I_b(x,y)
        \end{bmatrix}=\begin{bmatrix}
        f_r(\kappa _r(x,y))\\
        f_g(\kappa _g(x,y))\\
        f_b(\kappa _b(x,y))
        \end{bmatrix}.\\
    \end{split}
\end{equation}

Note that rendering linear data $\kappa$ into $I$ at an output color space (e.g., sRGB) takes a series of processing steps, which is always referred to as the ISP pipeline. It is therefore deficient to model such a process with the change of pixel intensity. To explain color preferably, the authors of \cite{RN61} propose a 24-parameter model with color transformation, which further extends the model in \eqref{gray_scale_calibration} to

\begin{equation}
    \label{color_calibration2}
    I\left ( x,y \right )=f\left (T \cdot \kappa \left ( x,y \right ) \right )=f\left (T \cdot\begin{bmatrix}
                 \kappa _{r}(x,y)\\
                 \kappa _{g}(x,y)\\
                 \kappa _{b}(x,y)
                    \end{bmatrix}
                    \right ).
\end{equation}
Here $T$ is a $3\times3$ matrix corresponding to the white balance step in ISP pipelines, $f(\cdot)$ is modeled with a $5$-th order polynomial for each channel and explained as color rendering function.  Inspired by \cite{RN61}, Kim et al. further found that the transformation in \eqref{color_calibration2} will fail for highly saturated colors. This is because the color mapping component including gamut mapping is missing in \eqref{color_calibration2}, and some supersaturated values outside the sRGB gamut cannot be converted to RAW accurately by this model\cite{RN62,RN63}. Based on these observations, the authors in \cite{RN62} and \cite{RN63} propose a model as
\begin{equation}
    \label{color_calibration3}
    \begin{split}
        I\left ( x,y \right )=f\left (T_s\cdot T_w \cdot\begin{bmatrix}
            \kappa _{r}(x,y)\\
            \kappa _{g}(x,y)\\
            \kappa _{b}(x,y)
               \end{bmatrix}
               \right ).
    \end{split}
\end{equation}
Here, $T_w$ is a $3\times3$ white balance diagonal matrix, and $T_s$ is another $3\times3$ matrix corresponding to color space transformation.

With the aforementioned reversible models, the rendered color image pixel intensities can be inverted back to their corresponding linear RAW values directly. However, there are some noticeable limitations in the aforementioned works:
\begin{itemize}
    \item As the authors suggested in \cite{RN61}, the color rendering function $f(\cdot)$ is mostly a scene-dependent process instead of a fixed property of a camera. Moreover, a fixed nonlinearity per channel/camera model as used in traditional radiometric calibration are often inadequate to model $f(\cdot)$.
    \item Work \cite{RN63} is an improved version of \cite{RN61}, but their experiments are carried out under stringent conditions, e.g., cameras should work on photographic reproduction mode to disable all the auto optimizers and use a fixed ISP pipeline, i.e., images for inverse transformation should be shot and processed under similar conditions to ensure the accuracy of the reverse conversion. Unfortunately, not all cameras support this mode, and it is hard to ensure the conditions.
    \item Models in \cite{RN61,RN62,RN63}  are established under the premise of a lot of simplifications. For example, a commonly used commercial ISP pipeline usually consist of many steps, but the pipeline in\cite{RN61,RN62,RN63} includes only a small number of them (e.g., white balance, gamut mapping, etc.), and others are modeled as noise processes.
\end{itemize}

\subsection{ISP Configuration \& In-Sensor Computing}
It has been mentioned in the Section~\ref{section:sec1} that conventional ISP pipeline is design for photography. For computer vision tasks, excessive processing may introduce cumulative errors and undermine the original information captured by image sensors \cite{RN14}. For example, demosaicing usually smooths the images, gamma compression corrects the inconsistency between the ways cameras capture scenes, monitors displays images and human visual system (HSV) processes light \cite{RN74}.

The optimal configuration for computer vision oriented ISP pipeline remains an open problem, where RAW dataset is crucial to solve this problem, since modern computer vision algorithms are mostly data-driven. In \cite{RN64}, the authors show both theoretically and experimentally that gradient-based features can be directly extracted from RAW Bayer images, such that the entire ISP pipeline can be bypassed completely. For deep learning based computer vision tasks, eight existing vision algorithms are extensively experimented in \cite{RN23} to find the minimal but sufficient ISP pipeline. The authors use the CRIP tool\cite{RN23} to generate RAW Bayer datasets by converting commonly used standard datasets such as MNIST\cite{RN76}, ImageNet\cite{RN75} and CIFAR-10\cite{RN77} back into RAW. The CRIP tool is implemented based on the model proposed in \cite{RN63}, but images in these datasets are from the Internet and captured using different cameras. Therefore, there may be a mismatch in camera models. Moreover, in CRIP, the noise needs to be explicitly added to the image signal in the reverse ISP pipeline. All these limitations may result in inaccurate estimation of RAW Bayer data, which may further affect the evaluation of stages in the ISP pipeline.

As mentioned in introduction, the gap between training set and testing set will decline the performance of in-sensor machine vision applications. \cite{RN70} proposes to solve this problem by finding a minimalistic ISP pipeline, consisting of gamma compression and pixel binning, to approximate the data distribution of images processed by ISP pipelines. Although the model's performance can be improved after these operations, several modifications need to be done to image sensors.

\begin{figure*}[htbp]
    \centering
    \includegraphics[width=7.3in]{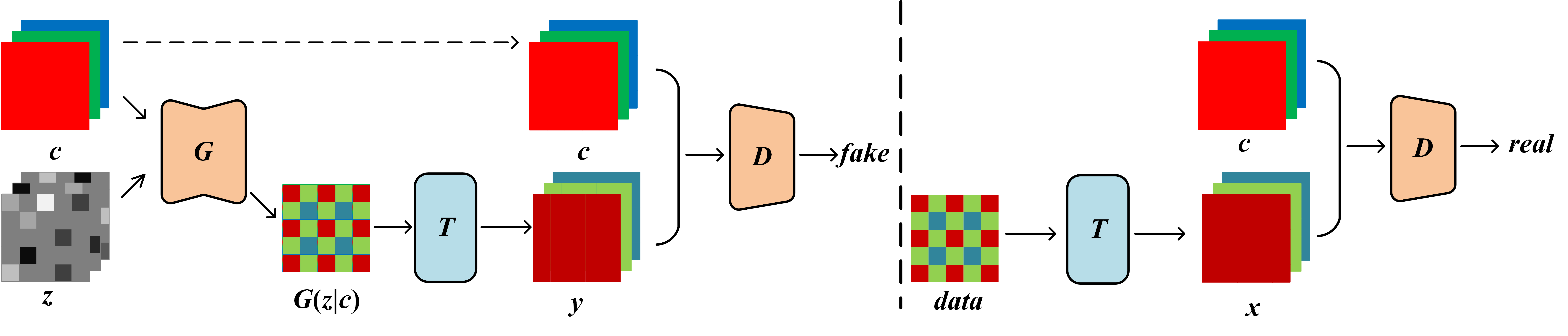}
    \caption{The training process of proposed method. The real and synthesized data are transformed by $T$ before fed into the discriminator.}
    \label{raw_gan_structure}
\end{figure*}

\subsection{GAN for Image-to-image Translation}
Image-to-image translation aims to convert an image from the source domain to a specified target domain, while preserving the main presentations of the input images. With the application of adversarial training, cGAN\cite{RN80} are increasingly used in image-to-image translation works with promising results \cite{RN66,RN67,RN68,RN69}.
Inspired by the success of GAN, we model the color-to-Bayer image conversion as sRGB domain to RAW domain translation, which can be effectively solved by GAN. The details of how to use GAN to generate high quality Bayer pattern images from their color versions will be discussed in the remaining of this paper.


\section{Proposed Method}
\subsection{Theoretical Analysis}

The basic idea is to model the color image to RAW Bayer pattern image transformation as a task of image-to-image translation, where the RAW Bayer pattern images and the color images are regarded as two different image styles.
The image-to-image transformation model used in this work is cGAN \cite{RN66}, whose objective can be expressed as
    \begin{equation}
        \label{valigan_loss_function}
        \begin{split}
            \underset{G}{min}\,\underset{D}max\,V(G,D)=&\mathbb{E}_{x\sim p_{data}}[\log(D(x|c))]\\+ &\mathbb{E}_{y\sim p_g}[\log(1-D(y))].
        \end{split}
    \end{equation}
Here, $p_{data}$ is the distribution of the real data, $p_g$ is the distribution of generated data which is learned by the generator $G$ through mapping prior noise $z \sim p_z$ to the data sample $G(z|c) \sim p_g$. Meanwhile, $x$ and $y=G(z|c)$ are the data sampled from $p_{data}$ and $p_g$, respectively, $c$ is the auxiliary information or the condition sampled from the target domain.



\begin{figure}
    \centering
    \includegraphics[width=3.5in]{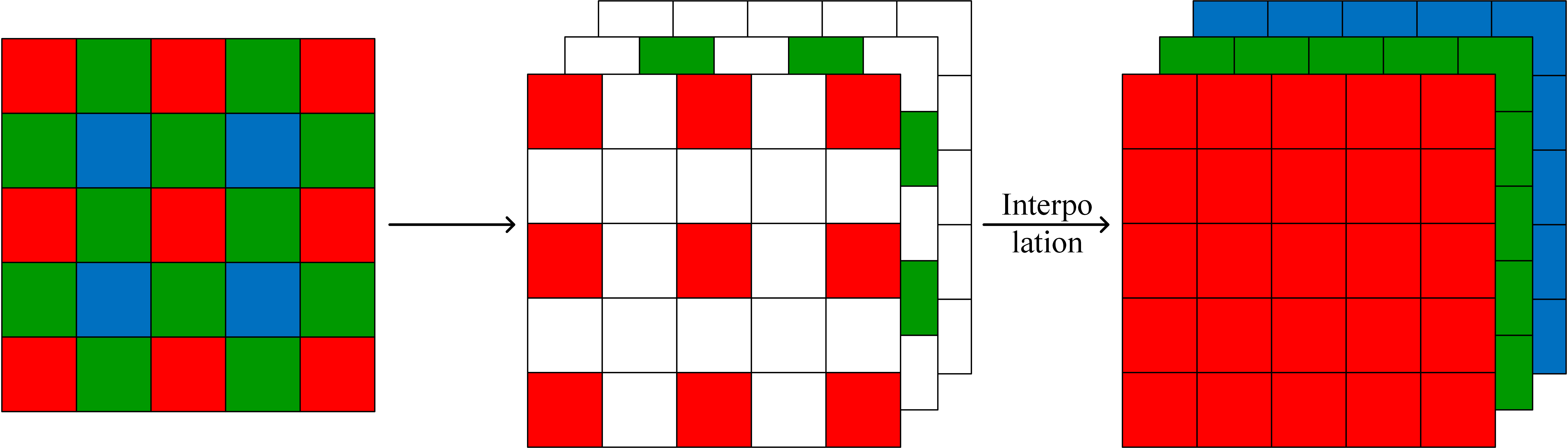}
    \caption{Illustration of demosaicing process.}
    \label{demosaicing}
\end{figure}



Suppose $T = \{T_1, T_2,...,T_m\}$ is a set of image transformation techniques, the real and generated data are transformed by the transformation $T(\cdot)$ before feeding into the discriminator $D_T$, then the min-max objective of cGAN would be
\begin{equation}
    \label{mygan_loss_function}
    \begin{split}
        \underset{G}{min}\,\underset{D_T}max\,V(G,D_T)=&\mathbb{E}_{x\sim p^T_{data}}[\log(D_T(x|c))]\\+ &\mathbb{E}_{y\sim p^T_{g}}[\log(1-D_T(y)],
    \end{split}
\end{equation}
where $p^T(\cdot)$ is the distribution of the corresponding transformed data. We will deduce the feasibility of using the transformed generated data $p^T_{g}$ to improve the learning of original data $p_{g}$ theoretically in the next part.

\begin{figure*}
    \centering
    \includegraphics[width=7.3in]{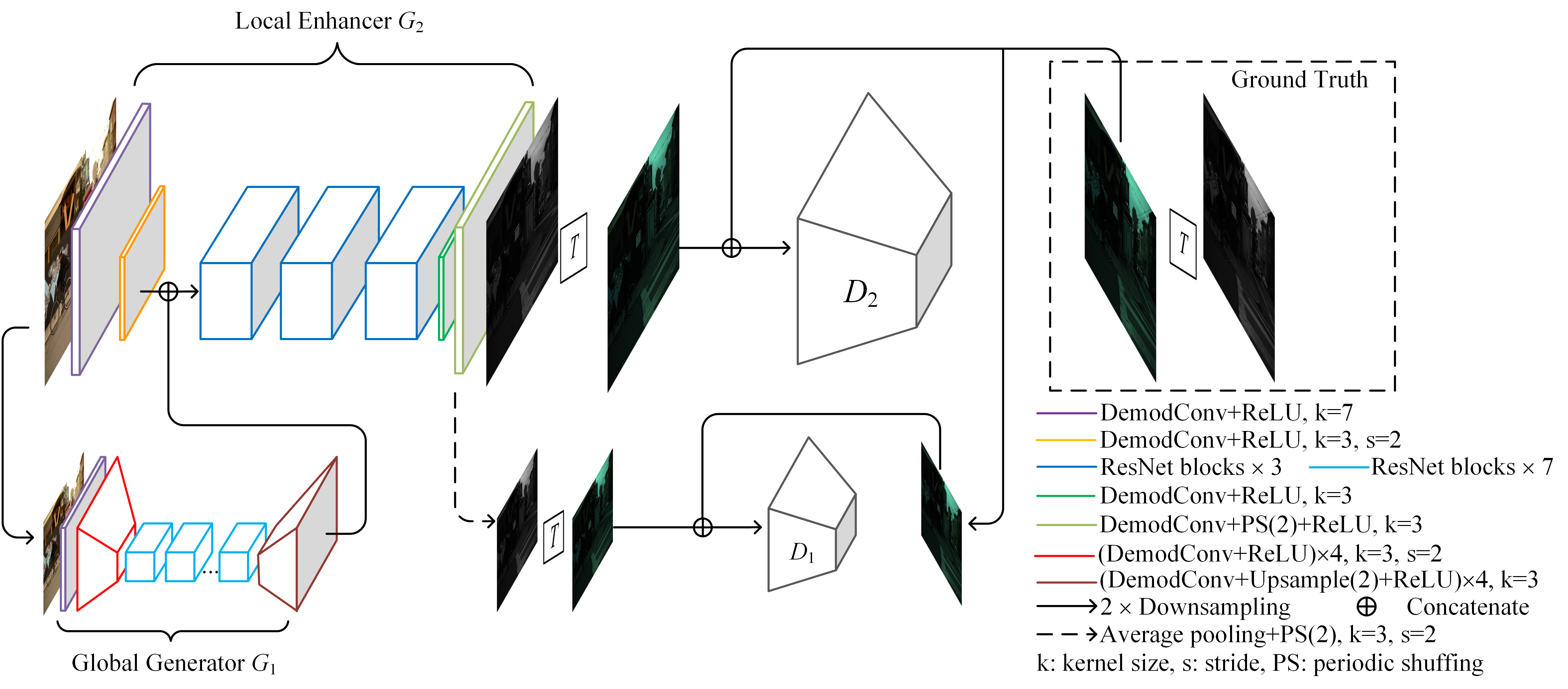}
    \caption{The overall architecture of the proposed method.}
    \label{architecture}
\end{figure*}




\subsubsection*{GAN Training with Transformed Data}
It has been shown in \cite{RN93} that suppose $T : \mathbb{X}  \rightarrow \mathbb{X}^T $ is a differentiable and invertible transformation which transforms $x$ to $x^T$, with fixed $G$, the optimal discriminator $D_{T}^{*}$ trained with transformed data $x^T$ can be expressed as
\begin{equation}
    \label{dt_and_d}
    \begin{split}
    D_{T}^{*}(x^T)&=\frac{p^T_{data}(x^T)}{p^T_g(x^T)+p^T_{data}(x^T)}\\ &=\frac{p_{data}(x)\left | \mathcal{J}^T(x) \right |^{-1}}{p_g(x) \left | \mathcal{J}^T(x) \right |^{-1}+p_{data}(x)\left | \mathcal{J}^T(x) \right |^{-1}}\\ &=\frac{p_{data}(x)}{p_g(x)+p_{data}(x)}\\&=D^{*}(x),
    \end{split}
\end{equation}
meaning that the optimal solution for $D^T$ and $D$ are exactly the same. Here, $\left | \mathcal{J}^T(x) \right|$ is the determinant of Jacobian matrix of transformation $T$. Therefore, the virtual training criterion\cite{RN80} of \eqref{mygan_loss_function} can be expressed as
\begin{equation}
    \label{gan_vitural}
    V(G,D_{T}^{*})=-\log(4)+2\cdot JS(p^T_{data}||p^{T}_{g}).
\end{equation}
Here, $JS(\cdot)$ is the Jensen–Shannon (JS) divergence between two distributions $p_{data}$ and $p_g$, which is non-negative. Therefore, the global minimal of \eqref{gan_vitural} is achieved when $JS(p^T_{data}||p^T_{g})=0$, and the only solution is $p^T_{data}=p^T_g$. According to $theorem 1$ in \cite{RN93}, under a differentiable and invertible transformation $T$, the JS divergence between
two distributions is invariant, which can be expressed as
\begin{equation}
    \label{JS_theorem}
    JS(p_{data}||p_{g})=JS(p^T_{data}||p^T_{g}).
\end{equation}
Combining \eqref{gan_vitural} and \eqref{JS_theorem}, we have
\begin{equation}
    \label{gan_vitural2}
    V(G,D^{*}_{T})=-log(4)+2\cdot JS(p_{data}||p_{g}).
\end{equation}
Equation \eqref{gan_vitural2} means that given an optimal discriminator $D^{*}_{T}$ trained with transformed samples, the objective of $V(G,D^{*}_{T})$ is to minimize the JS divergence between the distribution of generated data and original data, which guarantees $G$ to learn to generate original samples instead of transformed samples. 

Now, suppose the set $T$ contains only the demosaicing transformation, then the model would be as in Fig. \ref{raw_gan_structure}. Both the generated images and real RAW Bayer pattern images are demosaiced to color images before feeding into the discriminator $D_T$, which is equivalent to adding a constraint to the generated images, ensuring they can be demosaiced. That is, the generated images would maintain the Bayer pattern.

\subsubsection*{Demosaicing}
Fundamentally, demosaicing can be regarded as an interpolation operation. A typical demosaicing process is illustrated in Fig. \ref{demosaicing}. A Bayer pattern image is first split into three channels, i.e., R, G, and B, and the missing components are interpolated by using the information of existing components. There are different ways to interpolate these missing pixel components, which can be divided into two broad categories. The first one is to interpolate using only the pixels from the same channel. Typical methods in this category includes nearest neighbor interpolation, bilinear interpolation, cubic interpolation, etc. Another category is based on spectral correlation rules such as color ratio constancy or color difference constancy rules, during demosacing process. These methods take the strong spectral correlation between color components into account and generally obtain better performance\cite{RN1}.

For a specific Bayer pattern,  e.g., the $RGGB$ pattern as in Fig. \ref{demosaicing}, the demosaicing process can be described as
\begin{equation}
    \label{demosaic_eq}
    I_{k}=func[I_{bayer}\odot(\left \{P_{R},P_{G},P_{B}\right \}\otimes J)].
\end{equation}
Here, $\odot$ representes the Hadamard product operation, $\otimes$ is the Kronecker product operation, $func$ is a specific interpolation method for demosaicing and $J$ is an all-one matrix. For an $m\times n$ Bayer pattern image $I$, $J$ is a $\frac{m}{2} \times \frac{n}{2}$ matrix, $P_{k}$ is the $2\times 2$ Bayer pattern template, e.g., for $RGGB$ pattern,
$P_{R}=\begin{bmatrix}
        1 & 0\\
        0 & 0
        \end{bmatrix}$,
$P_{G}=\begin{bmatrix}
        0 & 1\\
        1 & 0
        \end{bmatrix}$,
$P_{B}=\begin{bmatrix}
        0 & 0\\
        0 & 1
        \end{bmatrix}$, $k\in \left \{ R,G,B \right \}$.
\vspace{0.1cm}

Correspondingly, for a demosaiced (without any further processing) three-channel color image, since the original captured pixels are not affected by interpolation, we can obtain them directly to restore the Bayer pattern image as
\begin{equation}
    \label{reverse_demosaicing}
    I_{bayer}=\sum (I_{k}\odot(P_{k}\otimes J)).
\end{equation}
This means that the demosaicing process is invertible.

Moreover, the Hadamard product $\odot$ and the Kronecker product $\otimes$ operations are differentiable. For interpolation operations, they only include simple four-arithmetic operations, which is also differentiable in the training process. 

\subsubsection*{Generate Bayer Pattern Images with Arbitrary Size}
Note that the RAW Bayer pattern images always have a large size. For example, images in the PASCALRAW dataset have a resolution of $4012\times 6034$. Training networks using images with such a high resolution will consume a lot of storage and computing resources, making them even not applicable for training. For many computer vision tasks, the size of input images are much smaller than that in the PASCALRAW dataset. For example, classification tasks usually take images with a resolution of $224\times 224$ \cite{RN81,RN82,RN83,RN84}. As for objection detection tasks, Faster R-CNN typically ingests $1000 \times 600$ images \cite{RN85}, single shot multibox detector (SSD) consumes $300\times 300$ or $512\times512$ images\cite{RN86}, RetinaNet takes $500\times500$ or $800\times 800$ input images \cite{RN90}, and YOLOv3 typically runs on either $320\times 320$ or $416\times 416$ or $608\times608$ images\cite{RN89}. However, down-sampling the Bayer pattern images directly will destroy the `mosaic' structure and result in the loss of color information\cite{RN64}. Although down-sampling with superpixel structure as proposed in \cite{RN64} can maintain the mosaic structure, the image quality may decrease obviously when the down-sampling factor is large.

One way to solve this problem is to split the images into small patches during the training process\cite{RN91}. However, since only local information can be learned, this solution is not friendly to model learning. The trained model may perform well on small patches, but there will be obvious stitches when stitching the small patches back into a complete image\cite{RN91}.

In this work, to generate images with arbitrary size, another transformation $T_2$ is added to the set $T$ to resize the demosacied image to a desired size. Since the resize operation is performed on demosaiced images, the previously mentioned problems on image quality and `mosaic' structure can be completely avoided.

\subsubsection*{Other Transformations}
Expect for above-mentioned transformations, other transformations or steps in ISP can also be added to set $T$. For example, the white balance stage, which multiple three parameters to $R$, $G$ and $B$ channel respectively, and the gamma compression is to compute the $a$ power of each pixel, where $a\in \left [ 0,1 \right ]$, both of them are invertible and differentiable transformations. Moreover, the proposed method is deduced from the GAN loss, which has the potential to be extended to other GAN models. In this paper, the proposed method is implemented on an improved pix2pixHD model to demonstrate the effectiveness of the proposed method. Detailed experimental results and corresponding discussions are presented in Section \ref{sec:image_quality}.


\subsection{Proposed Model Architecture}
In this work, cGAN is used as an example to demonstrate the proposed method. The overall architecture of the proposed model is illustrated in Fig. \ref{architecture}. Multi-scale generator and discriminator architectures are employed in the proposed model. The generator, consisting of two sub-networks, i.e., $G_1$ and $G_2$, generates images in a coarse-to-fine manner. The discriminators, consisting of $D_1$ and $D_2$, are  trained to discriminate real and synthesized images at 2 different scales. The inputs to $G$ are the finely rendered images, and the inputs to $G_1$ are $2\times$ down-sampled version of their $G_2$ counterparts. The inputs to $D$ are the concatenation of demosaiced version of $G$'s outputs and the corresponding ground truth. Both $G_1$ and $G_2$ consist of a convolution front-end, $n$ residual blocks, and a transposed convolution back-end.

The following modifications have been made to improve the quality of the synthesized images.
\begin{itemize}
    \item Water droplets artifacts elimination. In our initial design, we observed water droplets artifacts in the synthesized images as described in \cite{RN94}. Analysis results show that they are caused by instance normalization. Following the suggestion in \cite{RN94}, we replace the instance normalization with a weight demodulation operation as
    \begin{equation}
        \label{demodulate}
        w'_{ijk_hk_w}=\frac{s_i\cdot w_{ijk_hk_w}}{\sqrt{\sum _{i,k_h,k_w}(s_i\cdot w_{ijk_hk_w})^2+\varepsilon}},
    \end{equation}
    which is baked to a single convolution layer named demodulated convolution (DemodConv in Fig. \ref{architecture}). Here, $w$ is the original weight while $w'$ is the demodulated weights, $s_i$ is the scale factor corresponding to the $i$-th input feature map, $j$ is the $j$-th output feature map, $k_h$ and $k_w$ are the width and height of the kernel, respectively, and $\varepsilon$ is a small constant to avoid numerical issues.
    \item Checkboard artifacts alleviation. The checkboard artifacts can be observed when generating images by neural networks. This kind of artifacts is usually caused by deconvolution operation\cite{RN95}. To alleviate the checkboard artifacts, we replace the deconvolution layers with ``upsample$+$demodulated convolution'' operations, that is, the feature maps are first upsampled by nearest-neighbor, followed by a demodulate convolution operation \cite{RN110}. 

    \item To match the Bayer pattern. It has been mentioned in Section~\ref{section:sec1} that the Bayer pattern images consist of alternating $RGGB$ pixels. In order to matain the Bayer pattern, the last module of $G_2$ contains a "demodulated convolution $+$ periodic shuffing" operation. More specially, suppose there is a $3\times H\times W$ input color image and we want to generate an $H\times W$ Bayer pattern image. The output of $G_2$'s last convolution layer is a $4\times \frac{H}{2}\times\frac{W}{2}$ feature
    map, where each channel represents $RGGB$. Then a periodic shuffing operation \cite{RN96} is applied to generate an $H\times W$ image. Similarly, the input of $D_1$ is first average pooled to $4\times \frac{H}{4}\times\frac{W}{4}$, then a $\frac{H}{2}\times\frac{W}{2}$ image is generated by periodic shuffing.   
\end{itemize}

\subsection{The Overall Loss Function}
The overall loss function of the proposed model can be expressed as 
\begin{equation}
\label{overall loss function}
    \mathcal{L}=\mathcal{L}_{GAN}+\alpha_{1}\mathcal{L}_{FM}+\alpha_{2}\mathcal{L}_{VGG},
\end{equation}
where, $\mathcal{L}_{GAN}$ is the adversarial loss, $\mathcal{L}_{FM}$ is the feature matching loss, $\mathcal{L}_{VGG}$ is the perceptual loss and $\alpha_{1}$, $\alpha_{2}$ control the importance of $\mathcal{L}_{FM}$ and $\mathcal{L}_{VGG}$.
\begin{itemize}
    \item The adversarial loss is used to enable the model to generate data as close to the real data distributions as possible. With the two-scale discriminators $D_1$ and $D_2$ as shown in Fig. \ref{architecture}, the adversarial loss can be formulated as
        \begin{equation}
            \mathcal{L}_{GAN}=\underset{G}{min}\,\underset{D^1_T,D^2_T}{max}\,\sum_{k=1,2} V(G,D^k_T),
        \end{equation}
    where $V(G,D^k_T)$ is the single adversarial loss of the $k$-th discriminator $D_{k}$, as formulated in \eqref{mygan_loss_function}.

    \item The feature matching loss is used for learning information from multi-scale features by matching the intermediate feature maps of discriminator $D^k_T$ between the real and the synthesized images, which can be formulated as
    \begin{equation}
        \mathcal{L}_{FM}=\underset{G}{min}\,\sum_{k=1,2} l_{FM}(G,D^k_T).
    \end{equation}
    Here, $l_{FM}(G,D^k_T)$ is the feature matching loss from the $k$-th discriminator, given by
    \begin{equation}
        l_{FM}(G,D^k_T)=\mathbb{E}_{y\sim p^T_{g}}\sum_{i=1}^{M}\frac{1}{C_iW_iH_i}\left [ \left \| F_i(x)-F_i(y) \right \|_1 \right ],
    \end{equation}
    where $F_i(\cdot)$ is the output of $i$-th intermediate layer of discriminator $D_T^k$, $\|\cdot \|_1$ is the L1 distance, $M$ is the total number of layers and $H$, $W$, $C$ represents the height, width and channels of corresponding output feature maps, respectively.
    \item The perceptual loss is used for keeping the perceptual and semantic
    fidelity. Based on a pre-trained VGGNet \cite{RN109}, the perceptual loss is defined as
    \begin{equation}
        \mathcal{L}_{VGG}(x,y)=\underset{G}{min}\sum_{i=1}^{5}\frac{\lambda_i}{C_iW_iH_i}\left [ \left \| B_i(x)-B_i(y) \right \|_1 \right ].
    \end{equation}
    Here, $B_i$ is the $i$-th block of VGGNet and $\lambda_i$ is the corresponding weight coefficient of the $i$-th block.
\end{itemize}

\section{Experiments}
In this section, we implement the proposed method on an improved pix2pixHD model to convert the finely rendered color images to the corresponding Bayer pattern images. Experimental results on image quality as well as performance for computer vision tasks are presented to demonstrate the effectiveness of the proposed method. Datasets are first introduced in this section, followed by the details of the experiments setup and experimental results.
\subsection{Datasets}
There are two datasets used in our experiments, which are the Zurich RAW2RGB (ZRR) dataset \cite{RN100} and the PASCALRAW dataset \cite{RN55}. The ZRR dataset consists of RAW photos captured by a Huawei P20 mobile phone and a Canon 5D DSLR. These images are cropped into small patches. The images taken by Huawei P20 are used in our experiments because there are paired color and RAW images. The cropping procedure results in 90 thousand training patches and 2.4 thousand testing patches.
The PASCALRAW dataset, containing 4259 annotated RAW images and 3 annotated object classes (car, person and bicycle), is modeled following the manner of the PASCAL VOC dataset \cite{RN99}. All these images are captured by a Nikon D3200 DSLR camera, where both RAW images ($RGGB$ pattern) and the corresponding finely rendered color images are saved.

\subsection{Model Configuration}

In our training process, the initial learning rate is fixed to $2\times10^{-4}$ for discriminator and $2\times10^{-5}$ for generator. For optimization, an Adam optimizer with $\beta_1=0.5$, $\beta_2=0.999$ is used and the batch size is set to 4. The proposed model is implemented with PyTorch on four NVIDIA GeForce RTX 2080 TI GPUs and trained for more than $460$k steps.

\subsection{Image Quality Assessment Methods}
To evaluate the quality of synthesized Bayer pattern images, different image quality assessment methods are used in our experiments.
The Fréchet inception distance (FID) \cite{RN105}, which calculates the squared Wasserstein metric between two multidimensional Gaussian distributions, is a commonly used metric to evaluate the quality of images generated by GAN. Lower FID scores indicates higher similarity between two groups of images. The FID implementation in \cite{RN106} is adopted in our evaluation.
\begin{figure}[t]
    \centering
    \subfigure{
    \includegraphics[width=1.1in]{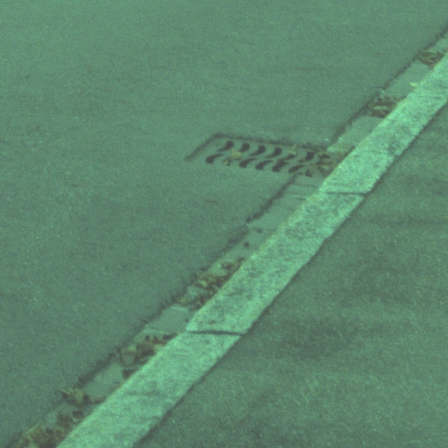}}
    \setcounter{subfigure}{0}
    \hspace*{-4.2mm}
    \subfigure[]{
    \label{fig:ZRDD_gt} 
    \includegraphics[width=1.1in]{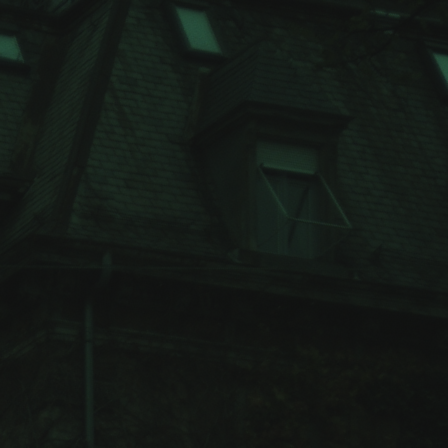}}
    \hspace*{-3mm}
    \subfigure{
    \includegraphics[width=1.1in]{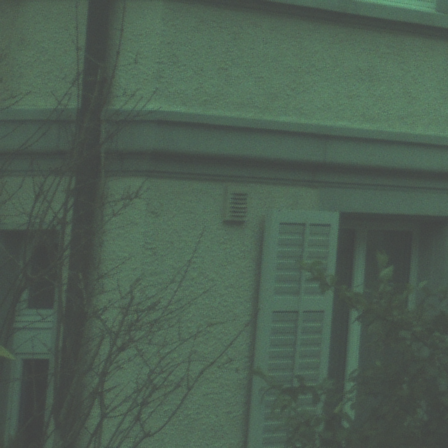}}
    \subfigure{
    \includegraphics[width=1.1in]{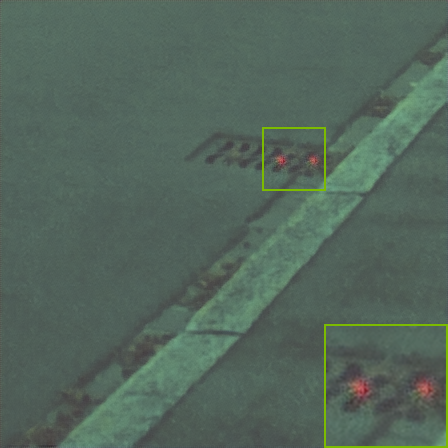}}
    \setcounter{subfigure}{1}
    \hspace*{-4.2mm}
    \subfigure[]{
    \label{fig:ZRDD_pix2pix} 
    \includegraphics[width=1.1in]{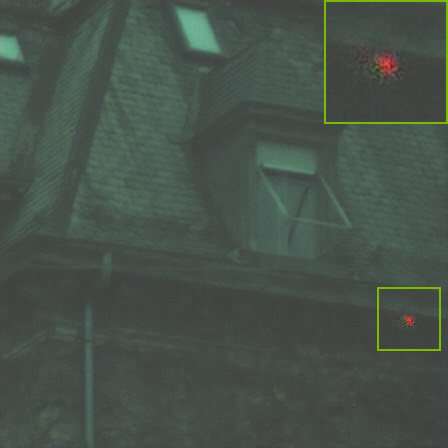}}
    \hspace*{-3mm}
    \subfigure{
    \includegraphics[width=1.1in]{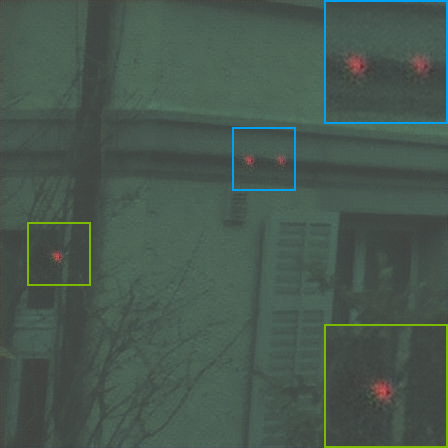}}
    \subfigure{
    \includegraphics[width=1.1in]{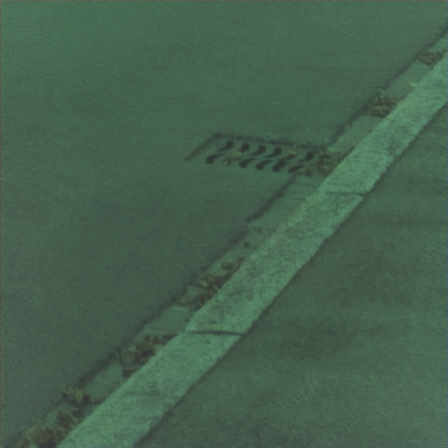}}
    \hspace*{-4.2mm}
    \setcounter{subfigure}{2}
    \subfigure[]{
    \label{fig:ZRDD_propose} 
    \includegraphics[width=1.1in]{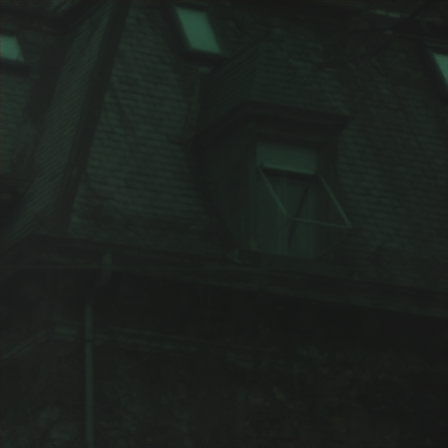}}
    \hspace*{-3mm}
    \subfigure{
    \includegraphics[width=1.1in]{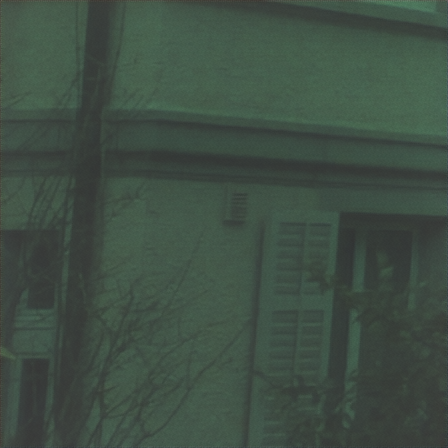}}

    \caption{Comparison between (a) original RAW Bayer images from ZRR dataset, (b) images generated by original pix2pixHD and (c) images generated by the proposed model. All these images are demosaiced using hybrid interpolation for display purpose.}
    \label{fig:compare_ZRDD}
\end{figure}

\begin{figure*}[t]
    \centering
    \subfigure[]{
    \label{fig:ZRDD_Examples_color} 
    \includegraphics[width=2.3in]{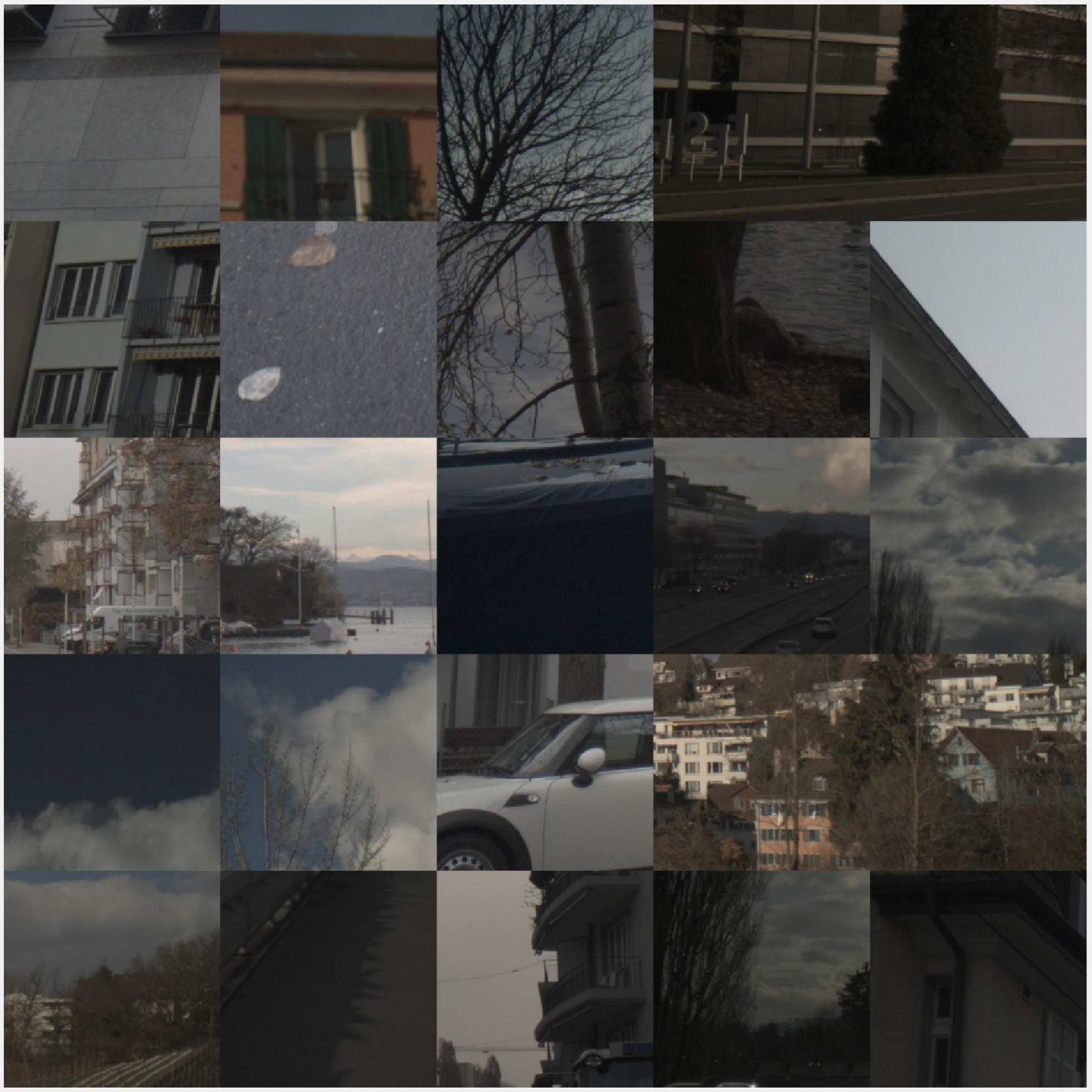}}
    \hspace{-3mm}
    \subfigure[]{
    \label{fig:ZRDD_Examples_bayer} 
    \includegraphics[width=2.3in]{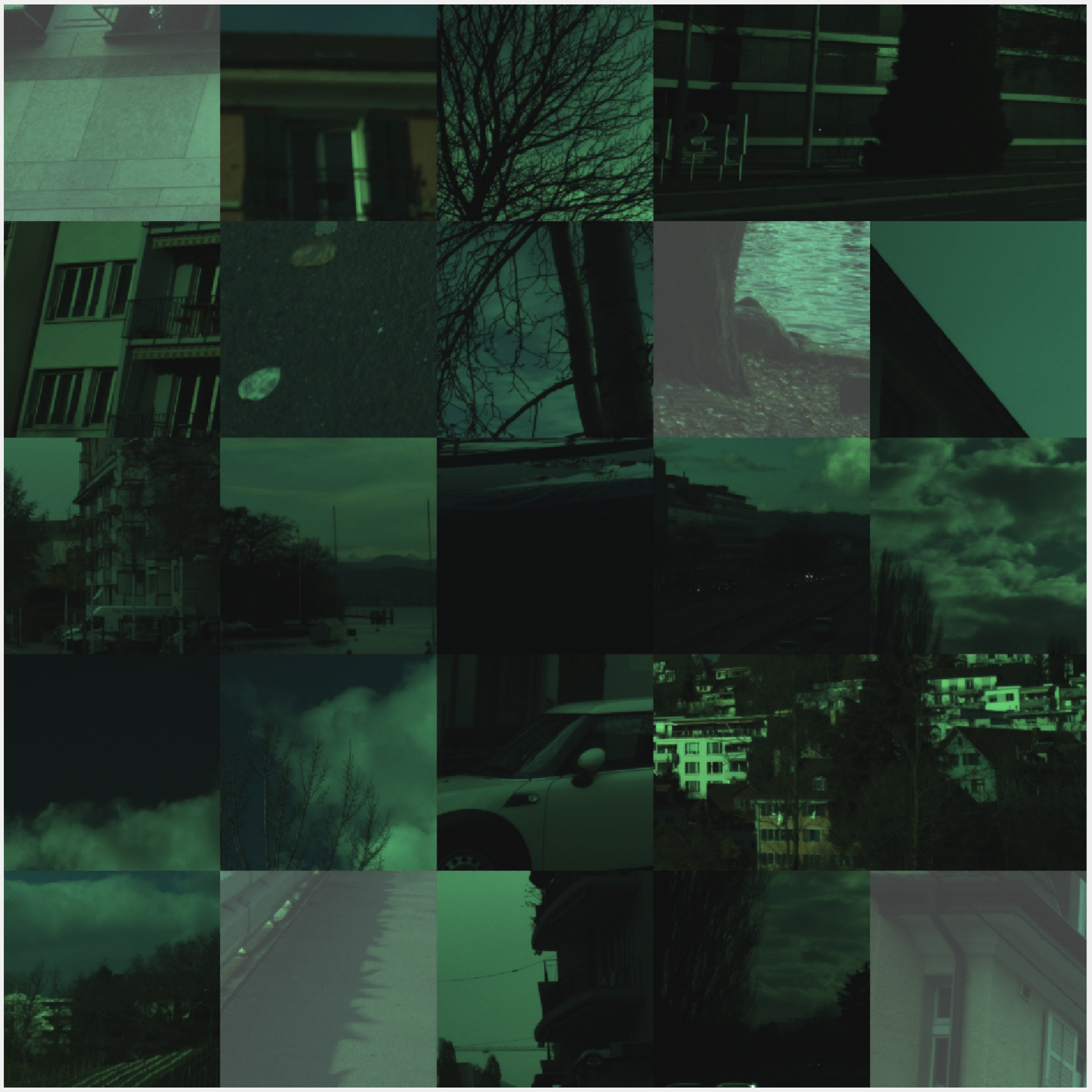}}
    \hspace{-3mm}
    \subfigure[]{
    \label{fig:ZRDD_Examples_syn} 
    \includegraphics[width=2.3in]{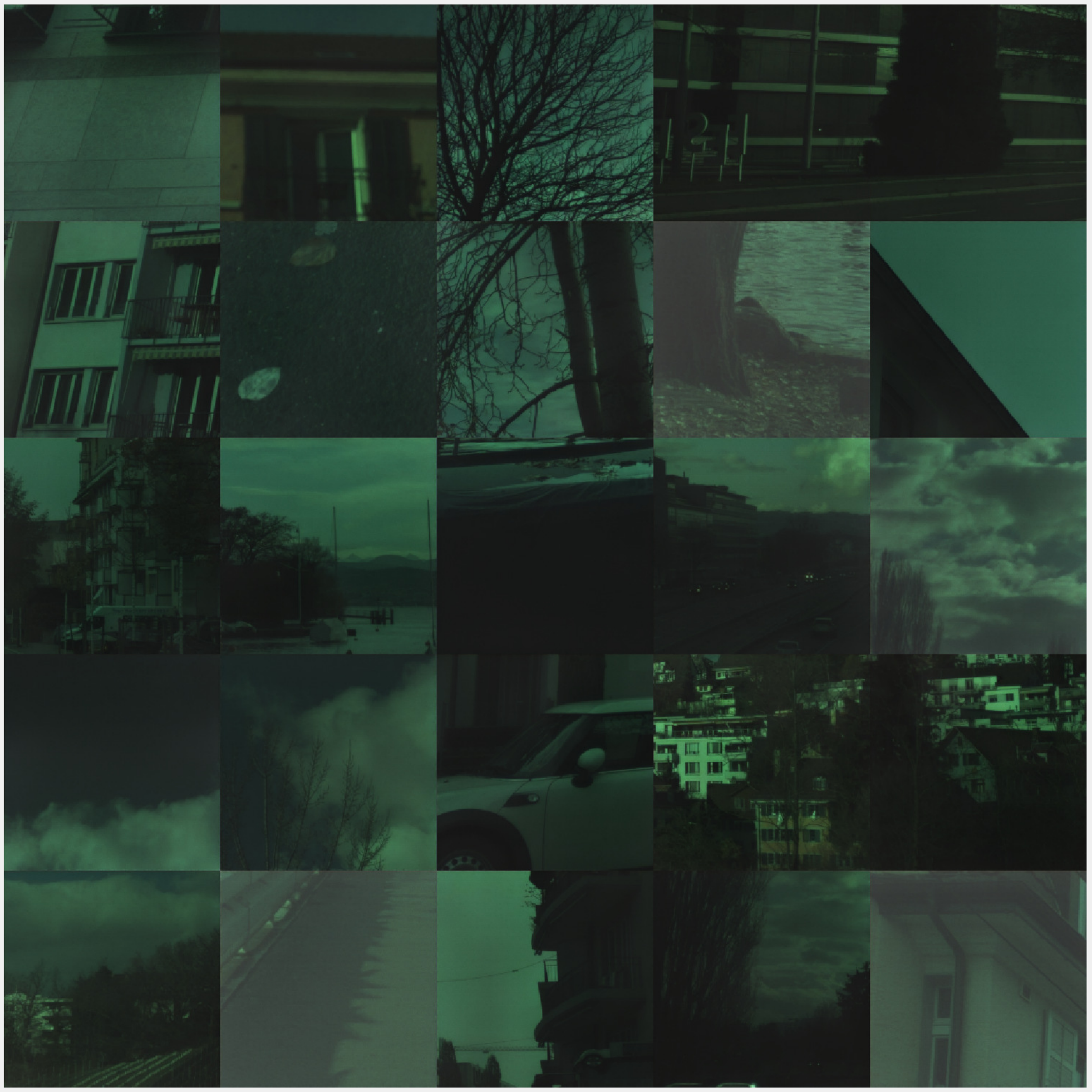}}
    \hspace{-3mm}
    \caption{Examples of image patches form the ZRR dataset. (a) The original color images. (b) The demosaiced original Bayer pattern images. (c) The demosaiced synthesized Bayer pattern images.}
    \label{fig:ZRDD_Examples}
    \end{figure*}

\begin{figure*}[t]
    \centering
    \subfigure[]{
    \label{fig:PASCAL_original_color} 
    \includegraphics[width=3.5in]{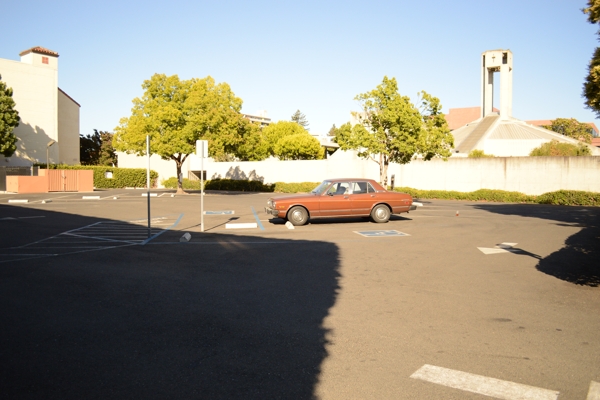}}
    \hspace{-3.5mm}
    \subfigure[]{
    \label{fig:PASCAL_gt_bayer} 
    \includegraphics[width=1.75in]{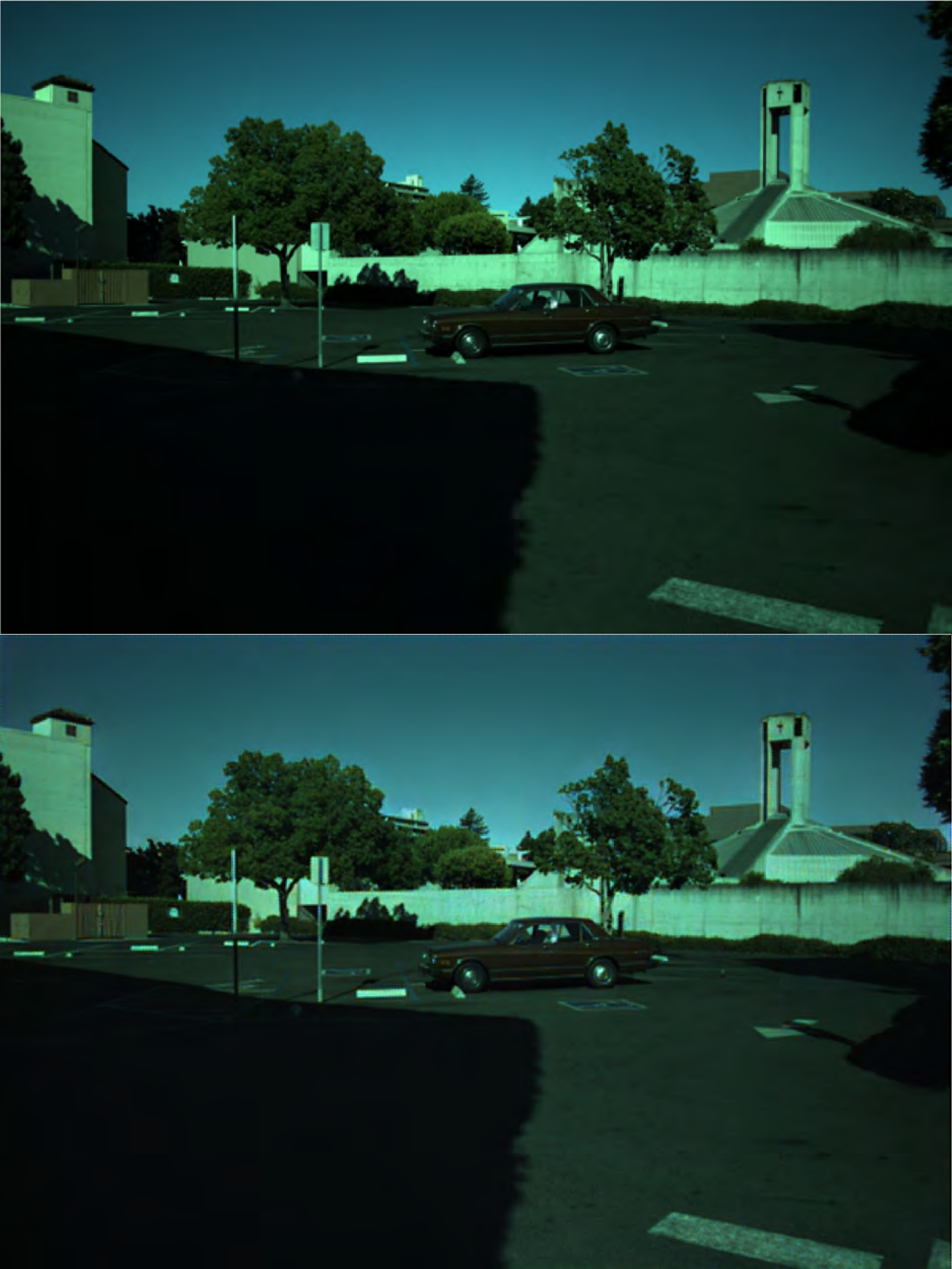}}
    \hspace{-3.5mm}
    \subfigure[]{
    \label{fig:PASCAL_gt_color} 
    \includegraphics[width=1.75in]{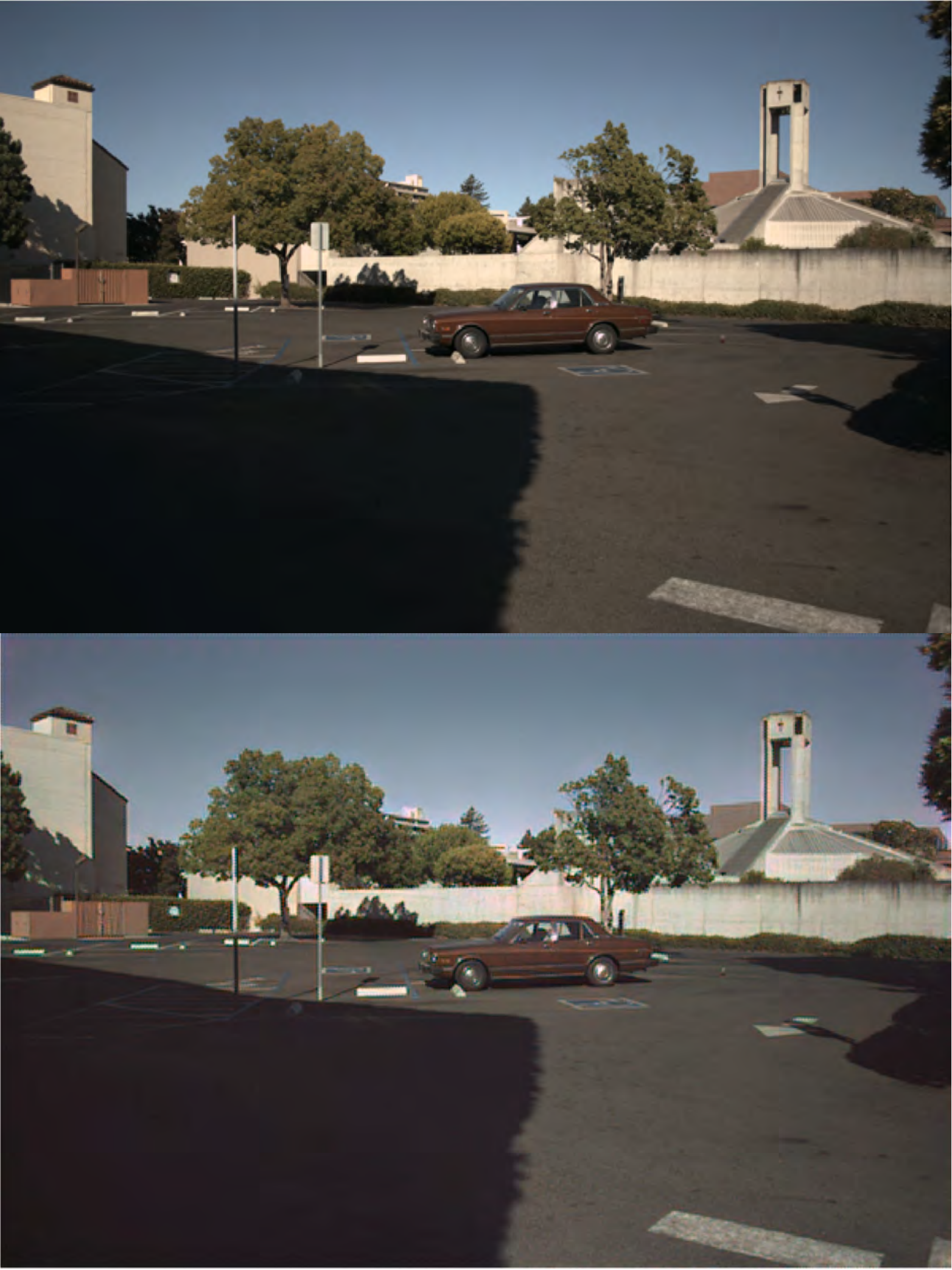}}

    \caption{Visual comparison of images processed by different pipelines. (a) An original color image from the PASCALRAW dataset. (b) The demosaiced version of the corresponding original (top) and synthesized (bottom) Bayer pattern images. (c) The fully processed original (top) and synthesized (bottom) Bayer pattern images by a simple ISP pipeline.}
    \label{fig:compare_PASCAL}
    \end{figure*}

    \begin{figure*}[t]
        \centering
        \subfigure[]{
        \label{fig:pascal_Examples:color} 
        \includegraphics[width=2.3in]{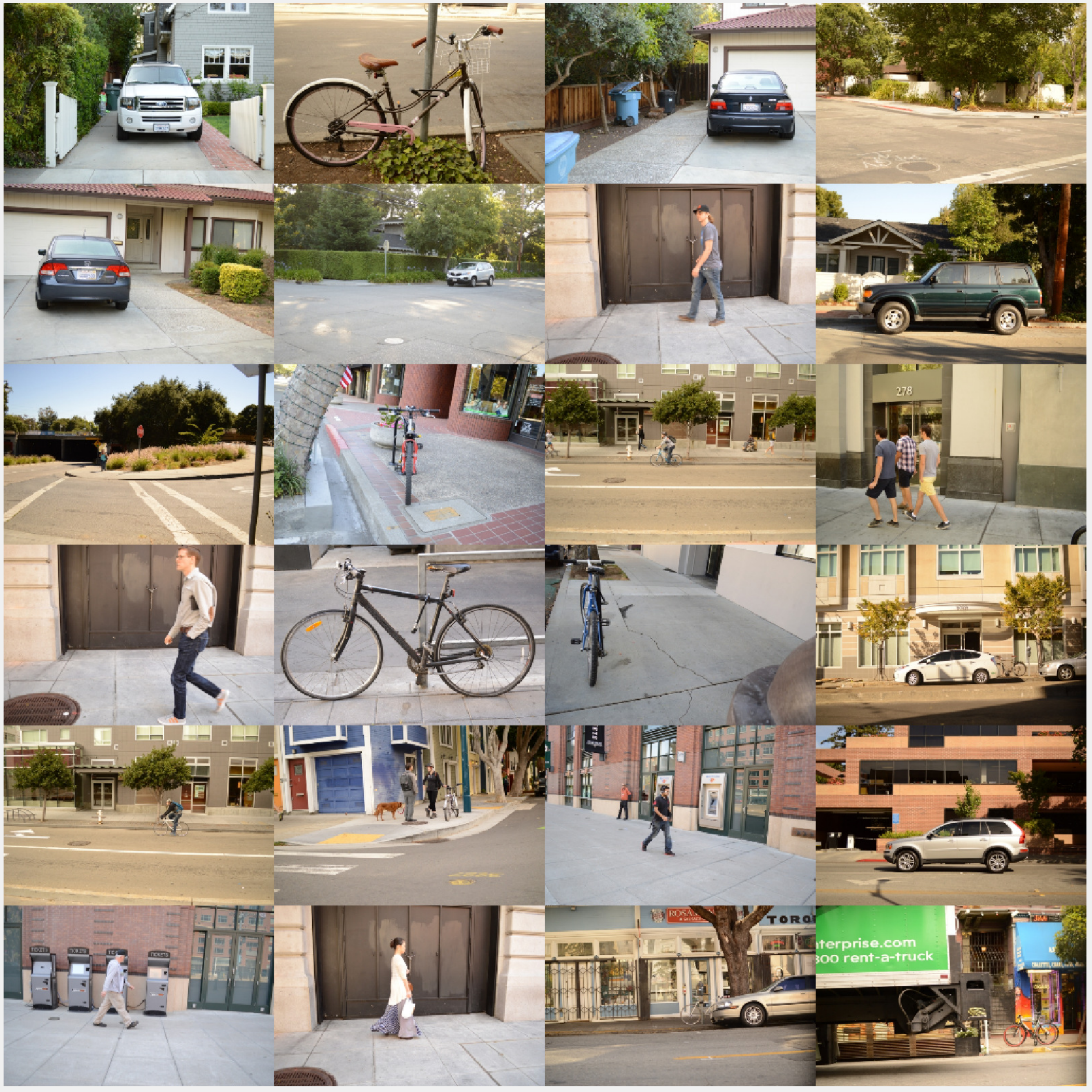}}
        \hspace{-3mm}
        \subfigure[]{
        \label{fig:pascal_Examples:bayer} 
        \includegraphics[width=2.3in]{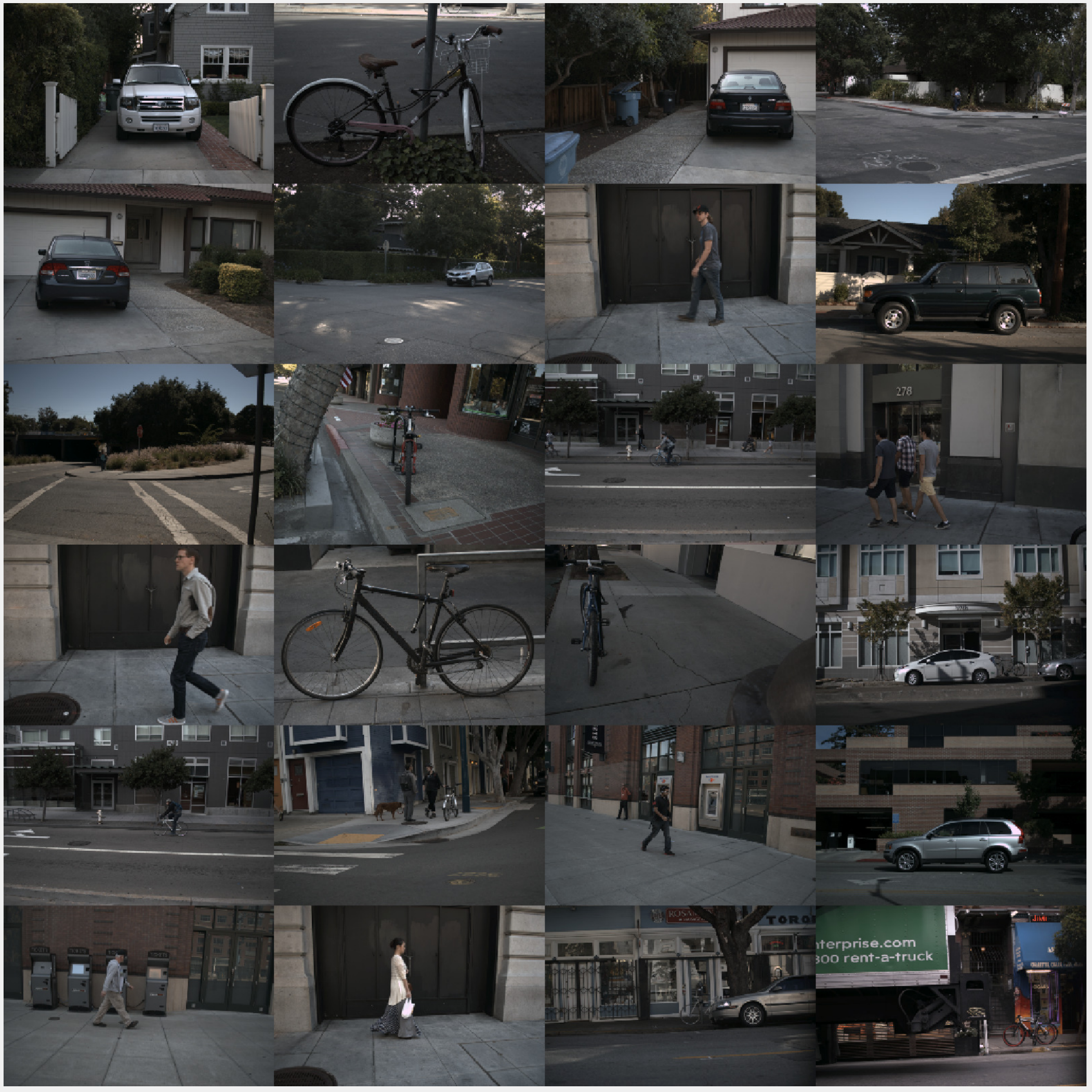}}
        \hspace{-3mm}
        \subfigure[]{
        \label{fig:pascal_Examples:syn} 
        \includegraphics[width=2.3in]{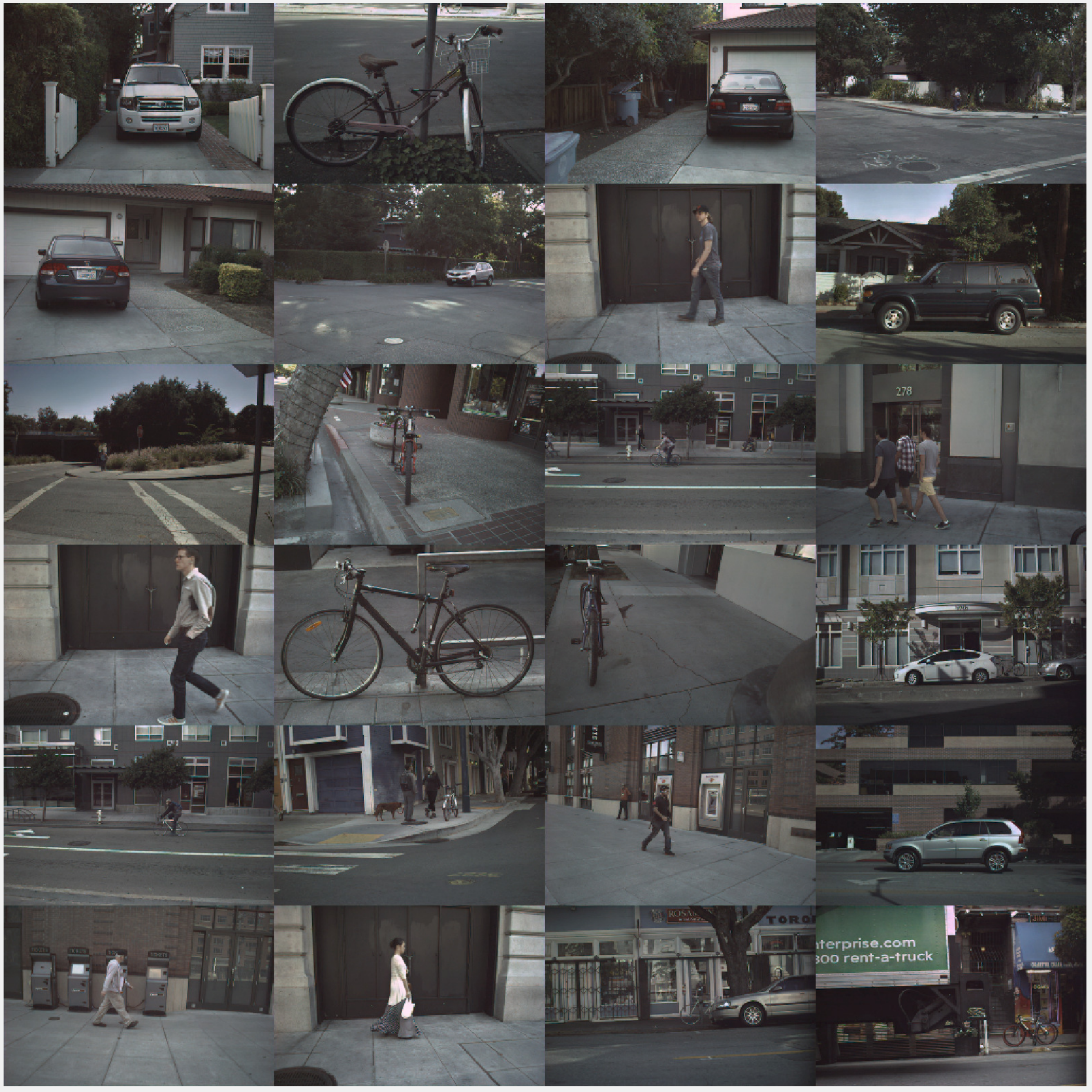}}
        \hspace{-3mm}
        \caption{Examples of original and synthesized images form the PASCALRAW dataset. (a) The original color images. (b) The fully processed (using a simple ISP pipeline) original Bayer pattern images. (c) The fully processed synthesized Bayer pattern images.}
        \label{fig:pascal_Examples}
        \end{figure*}

The peak signal to noise ratio (PSNR) and mean structural similarity (MSSIM) \cite{RN16} are two commonly used full-reference image quality metrics. For two given images, the PSNR can be calculated as
\begin{equation}
    \label{psnr}
    PSNR=10\log_{10}\left(\frac{(2^b-1)^2}{MSE}\right).
\end{equation}
Here, $b$ is the bit depth of images, and mean squared error (MSE) is defined as
\begin{equation}
    \label{mse}
    MSE=\frac{1}{H\times W}\sum ^H_{h=1} \sum ^W_{w=1}(I_1(h,w)-I_2(h,w)),
\end{equation}
where $H$ and $W$ are the height and width of images, respectively.
To maintain the consistency with MSE metric, in our experiments, the average PSNR among a specific dataset is calculated from average MSE as
\begin{equation}
    \label{ave_psnr}
    AVE\_PSNR=10\log_{10}\left(\frac{(2^b-1)^2}{\frac{1}{n}\sum_{i=1}^{n}MSE_i}\right),
\end{equation}
where $n$ is the number of images in the corresponding dataset.

MSSIM compares the luminance, contrast and structure of two images. To compute MSSIM, the images are splited into blocks and the average SSIM among all block pairs is calculated. Note that MSSIM ranges from 0 to 1, where 1 means two images are identical.




\begin{table}[]
    \centering
    \caption{Comparison Results between The Demosaiced Real Bayer Pattern Images and Demosaiced Synthesized Bayer Pattern Images}
    \label{tab:compare_eresults_ZRDD}
    \setlength{\tabcolsep}{10pt}{
    \begin{tabular}{lccc}
    \toprule
                        &\textbf{FID}    &\textbf{PSNR}   &\textbf{MSSIM}\\
    \midrule
    Original pix2pixHD    &22.87         &18.45  &0.86\\
    Proposed              &13.38         &20.17   &0.88\\
    \bottomrule
    \end{tabular}
    }
\end{table}

\subsection{Image Quality Comparison}
\label{sec:image_quality}
We implement the proposed model and  train it on the ZRR dataset. To better demonstrate the effectiveness of the proposed architecture, we also train an original pix2pixHD framework \cite{RN69} with same settings. The trained models are tested on ZRR testing set (2462 pairs) for image quality evaluation. Note that we use hybrid interpolation \cite{RN108} for all the transformation $T$ in our model. 

Table \ref{tab:compare_eresults_ZRDD} presents the comparison results between the original Bayer pattern images and synthesized Bayer pattern images. Both images are demosaiced using hybrid interpolation, and the FID, PSNR and MSSIM metrics are calculated with demosaiced ground truth, Bayer pattern images (the original RAW Bayer images from the cameras). It can be found that for all these metrics, the proposed model outperforms the original pix2pixHD model. To have a visual comparison, Fig. \ref{fig:compare_ZRDD} shows the demosaicd versions of ground truth, images generated by the original pix2pixHD and images generated by the proposed method. It can be found from Fig. \ref{fig:ZRDD_pix2pix} that the original pix2pixHD model may generate random artifacts (highlighted by boxes) in the training process and hard to be get rid of. Moreover, there will be obvious differences in the color saturation and overall brightness between the images generated by the original pix2pixHD model and the ground truth, while images generated by the proposed model is closer to the ground truth visually. Fig. \ref{fig:ZRDD_Examples} gives more examples from the ZRR dataset.

As mentioned before, the proposed model is able to generate Bayer pattern images with arbitrary size. To achieve this, another transformation $T_2$ needs to be added after the demosaicing step to resize the image to the desired size. As we have discussed in the Introduction, directly change the size of Bayer pattern images will cause image quality degradation and loss of color information. The proposed method avoids these problems. As a visual comparison, Fig. \ref{fig:compare_PASCAL} shows (a) the original color image, (b) the demosaiced (using hybrid interpolation) original (real) and synthesized Bayer pattern images, (c) the fully processed original and syhthesized Bayer pattern images. Note that the images in Fig. \ref{fig:PASCAL_gt_color} are processed by a simple ISP pipeline consisting of linearization, denoise, demosaicing, white balance and gamma compression. To compute the indices, the real and synthesized Bayer pattern images are demosaiced using hybrid interpolation. The average FID, PSNR and MSSIM calculated between demosaiced real Bayer pattern images and demosaiced synthesized Bayer pattern images are 6.956, 29.061 and 0.887, respectively, indicating the synthesized images have a good quality and very close to real images visually.

It can also be found that there exists difference between the Bayer pattern images processed by different ISP pipelines. Take white balance as an example, Fig. \ref{fig:PASCAL_original_color} is processed by the camera's white balance algorithm, while images in Fig. \ref{fig:PASCAL_gt_color} are processed by the gray world algorithm \cite{RN107}. Fig. \ref{fig:pascal_Examples} shows more examples of images from the PASCALRAW dataset. As can be seen,  mapping a rendered color images to their corresponding Bayer domain can also be used for re-rendering images in photography.

\begin{figure}[t]
    \centering

    \subfigure[]{
    \label{fig:boxes:gt} 
    \includegraphics[width=1.7in]{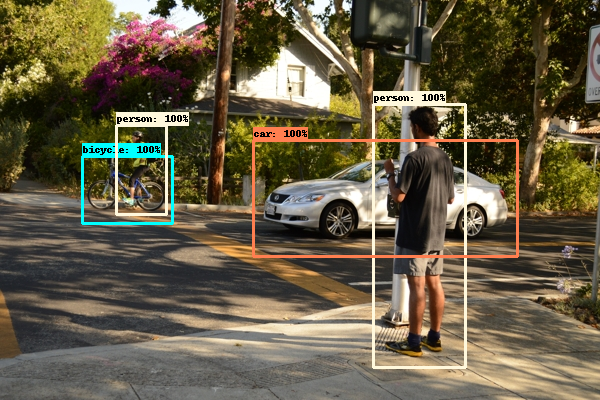}}
    \hspace{-3mm}
    \subfigure[]{
    \label{fig:boxes:color} 
    \includegraphics[width=1.7in]{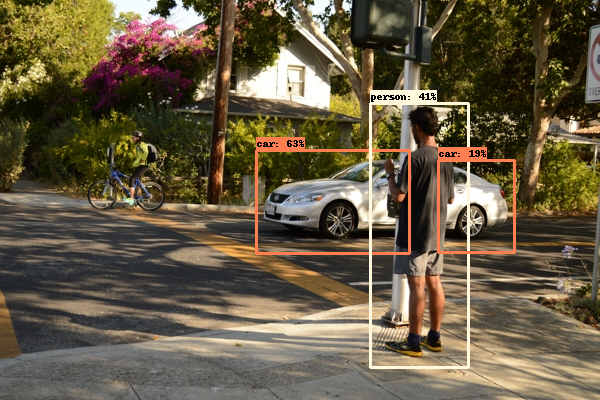}}
    \hspace{-3mm}
    \subfigure[]{
    \label{fig:boxes:bayer} 
    \includegraphics[width=1.7in]{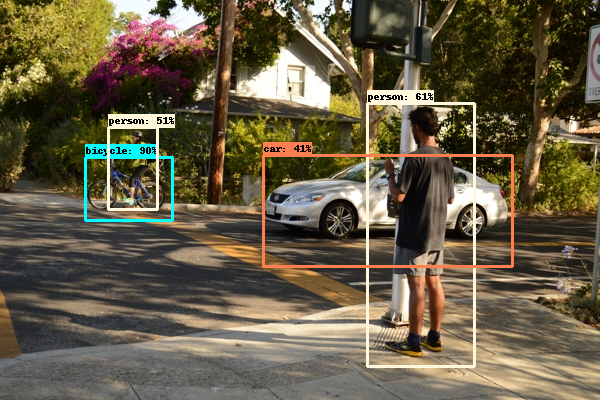}}
    \hspace{-3mm}
    \subfigure[]{
    \label{fig:boxes:syn} 
    \includegraphics[width=1.7in]{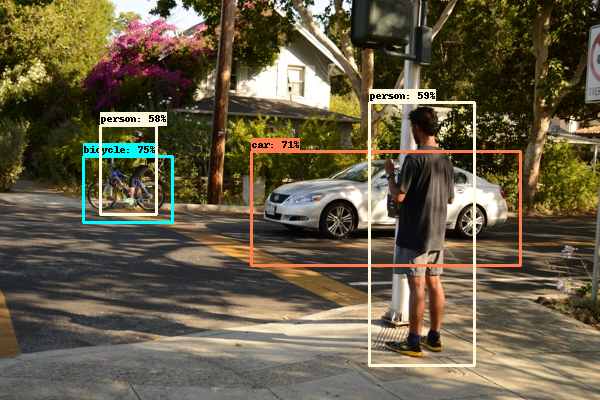}}
    \hspace{-3mm}
    \caption{Detected bounding boxes when apply different models on original Bayer pattern images. (a) The ground truth. (b) The model trained on the color images. (c) The model trained on the original Bayer pattern images. (d) The model trained on the synthesized Bayer pattern images.}
    \label{fig:boxes}
    \end{figure}

\subsection{Bayer Pattern Images for End-to-end Computer Vision}
As mentioned before, there are gaps between training data and testing data for end-to-end (from RAW Bayer image to computer vision results) computer vision architectures. Nowadays, almost all the computer vision models are trained on finely rendered color images, such that directly apply these models to Bayer pattern images will cause significant performance degradation. Since the synthesized RAW Bayer pattern images and finely rendered color images correspond to the same senses, we propose to use the synthesized Bayer pattern images generated by our method to train or fine-tune the computer vision models to improve their performance on Bayer pattern images, enabling end-to-end (from Raw Bayer images to vision tasks) computer vision system design. Note that in end-to-end vision systems, since the entire ISP pipeline is skipped, the overall system complexity will be reduced significantly \cite{RN64}, leading to reduced power consumption and more compact system design. Moreover, by skipping the ISP pipeline, the  latency of the entire vision system, which is crucial in applications such as autonomous vehicles, can also be significantly reduced.

In this experiment, the object detection task is taken as an example, where Faster R-CNN, SSD and YOLO-v3 frameworks are used as detection backbones. Since the PASCALRAW dataset provides both RAW Bayer pattern images and color images, we convert the rendered color images in the PASCALRAW dataset to their Bayer pattern counterparts, and train models on color images, original Bayer pattern images and synthesized Bayer pattern images to evaluate their performance.  

Table \ref{tab:object_detection_corresponding} presents the object detection results, where training and evaluation are performed on same version of datasets, e.g., both training and testing are performed on the color image dataset. It can be found that for all these frameworks, on the strict testing standards, i.e., AP (the Intersection-over-Union (IoU) from 0.5:0.95) and AP\textsubscript{75} (IoU=0.75), the detection performance on finely rendered color images is slightly better (about $2\%\sim5\%$) than that on original Bayer and synthesized Bayer datasets. But these difference will be further reduced for AP\textsubscript{50} (IoU=0.5). This indicates that some stages in the ISP pipeline may be beneficial for object detection tasks, or more likely the models are designed for finely rendered images without taking the properties of Bayer pattern images, such as alternating RGB pattern, into account. Moreover, the detection rate on original Bayer dataset is very close to that on synthesized Bayer image dataset for all these testing standards.

Table \ref{tab:object_detection_onbayer} presents the object detection results where the models are trained on color, original Bayer and synthesized Bayer datasets, and evaluations are performed on original Bayer pattern images. As can be seen from row 2, 5 and 8 of Table \ref{tab:object_detection_corresponding} and Table \ref{tab:object_detection_onbayer}, the average precision of object detection decreases significantly when apply the model trained on color dataset to original Bayer pattern images directly. The performance degradation can be as high as 10\% for the strict testing standards. But if we train the model on synthesized Bayer images, and apply it on original Bayer pattern images, the object detection performance can be improved significantly, which is very close the model trained on original Bayer pattern images. Moreover, it can be found from Table \ref{tab:object_detection_corresponding} and Table \ref{tab:object_detection_onbayer} (rows 4, 7 and 10) that if the model is trained on synthesized Bayer pattern image dataset, there is little difference on performance when evaluated on synthesized or original Bayer pattern image dataset.

Fig. \ref{fig:boxes} gives the ground truth (Fig. \ref{fig:boxes:gt}) and detection performance when apply the models trained on different datasets to original Bayer pattern images (Fig. \ref{fig:boxes:color}-\ref{fig:boxes:syn}). Note that detected objects (the bounding boxes) are drawn on the finely rendered color images for the convenience of visuality. It can be found from Fig. \ref{fig:boxes:color} that there are detection failures when apply the model trained on the color images to the original Bayer pattern images directly, which are also the main cause of the performance degradation presented in Table \ref{tab:object_detection_onbayer}. One of the problem is the miss detection, especially for small objects or objects in the dark scenes. For example, the person who ride a bicycle on the left upper corner in Fig. \ref{fig:boxes:color}. Another problem is that the detector can only recognize part of the objects, leading to the bounding boxes cannot cover the objects completely (the car in Fig. \ref{fig:boxes:color}). In this case, although the objects are detected, the bounding boxes will be regarded as false positive since the IoU score between the bounding boxes and the ground truth does not meet the threshold. Moreover, for the models trained on the original Bayer pattern images and synthesized Bayer pattern images in Fig. \ref{fig:boxes:bayer} and \ref{fig:boxes:syn}, the detection results are close to each other, except for a slight difference in confidence score and the boundary bounding boxes, which can be judged as true positive.     


\begin{table}
    \caption{Object Detection Results When Training and Evaluation are on the Same Version of Dataset}
    \label{tab:object_detection_corresponding}
    \centering
    \setlength\arrayrulewidth{0.3pt}
    \renewcommand{\arraystretch}{1.2}
    \setlength{\tabcolsep}{10pt}{
    \begin{tabular}{ccccc}
    \toprule
    \multicolumn{1}{l}{}         & Training Set & AP & AP\textsubscript{50} & AP\textsubscript{75}  \\
    \cline{2-5}
    \multirow{3}{*}{Faster R-CNN} & Color        & 71.94                 & 96.54                & 87.17                 \\
                                 & Bayer        & 69.56                 & 95.46                & 84.21                 \\
                                 & Synthesized  & 69.19                 & 95.13                & 83.52                 \\
    \cline{2-5}
    \multirow{3}{*}{SSD300}      & Color        & 66.48                 & 94.59                & 79.69                 \\
                                 & Bayer        & 63.91                 & 92.90                & 75.49                 \\
                                 & Synthesized  & 63.33                 & 92.33                & 74.74                 \\
    \cline{2-5}
    \multirow{3}{*}{Yolo-v3}     & Color        & 73.32                 & 96.21                & 88.25                 \\
                                 & Bayer        & 71.29                 & 95.37                & 85.31                 \\
                                 & Synthesized  & 70.86                 & 95.41                & 85.20                 \\
    \bottomrule
    \end{tabular}}
    \end{table}

\begin{table}
    \caption{Object Detection Results When Train Models on Different Datasets and Perform Evaluation on Real Bayer Dataset}
    \label{tab:object_detection_onbayer}
    \centering
    \setlength\arrayrulewidth{0.3pt}
    \renewcommand{\arraystretch}{1.2}
    \setlength{\tabcolsep}{10pt}{
    \begin{tabular}{ccccc}
    \toprule
    \multicolumn{1}{l}{}         & Training Set & AP & AP\textsubscript{50} & AP\textsubscript{75}  \\
    \cline{2-5}
    \multirow{3}{*}{Faster R-CNN} & Color        & 62.09                   & 91.26                  & 74.67                   \\
                                 & Bayer        & 69.56                   & 95.46                  & 84.21                   \\
                                 & Synthesized  & 68.81                   & 94.83                  & 82.95                   \\
    \cline{2-5}
    \multirow{3}{*}{SSD300}      & Color        & 57.45                   & 86.85                  & 67.14                   \\
                                 & Bayer        & 63.91                   & 92.90                  & 75.49                   \\
                                 & Synthesized  & 63.21                   & 92.58                  & 73.94                   \\
    \cline{2-5}
    \multirow{3}{*}{Yolo-v3}     & Color        & 64.54                   & 90.93                  & 75.76                   \\
                                 & Bayer        & 71.29                   & 95.37                  & 85.31                   \\
                                 & Synthesized  & 70.82                   & 95.31                  & 85.20                   \\
    \bottomrule
    \end{tabular}}
    \end{table}

\section{Conclusion}
 In this paper, we propose a method to add constraints that are un-formulatable in GAN training. By using the proposed method, we implemented a model to synthesize RAW Bayer pattern images with arbitrary size from their color image counterparts. Based on the invariant of JS divergence between two distributions, we first theoretically deduce the feasibility of using transformed generated data to improve the learning of the generator on original data. Then the proposed architecture is implemented, along with some well-designed modules, which match the structure of Bayer pattern to improve the quality of synthesized images. Experimental results show that under the same training condition, the proposed method is able to generate RAW Bayer pattern images with better FID score, PSNR and MSSIM than existing methods, and the training process is more stable. The proposed method can be used to generate image dataset for applications which need RAW Bayer pattern images. For example, works aiming to explore the optimal configuration of ISP pipeline for CV performance (instead of for photography), works in the field of in-sensor computing/smart image sensor design, or even photography. It is also shown in the experiments that by training object detection frameworks using the synthesized RAW Bayer images, they can work in an end-to-end manner with negligible performance degradation, i.e., RAW Bayer images (from the image sensor) can be directly fed to object detection algorithms without image processing. By skipping the ISP pipeline, both computational complexity and power consumption of the overall vision system can be significantly reduced. Moreover, without an image processor in the middle, image sensors can be directly integrated with back-end vision processors, making in-sensor computing a practical approach.


\bibliographystyle{ieeetr}
\bibliography{High_quality_raw_bayer_pattern_images}

\begin{thebibliography}{10}

\bibitem{RN4}
R.~Ramanath, W.~E. Snyder, Y.~Yoo, and M.~S. Drew, ``Color image processing
  pipeline,'' {\em IEEE Signal Processing Magazine}, vol.~22, no.~1,
  pp.~34--43, 2005.

\bibitem{RN14}
F.~Heide, M.~Steinberger, Y.~T. Tsai, M.~Rouf, D.~Paj{\k{a}}k, D.~Reddy,
  O.~Gallo, J.~Liu, W.~Heidrich, K.~Egiazarian, {\em et~al.}, ``Flexisp: A
  flexible camera image processing framework,'' {\em ACM Transactions on
  Graphics}, vol.~33, no.~6, pp.~231:1--231:13, 2014.

\bibitem{RN61}
A.~Chakrabarti, D.~Scharstein, and T.~E. Zickler, ``An empirical camera model
  for internet color vision,'' in {\em Proc. British Machine Vision
  Conference}, vol.~1, p.~4, Sept. 2009.

\bibitem{RN62}
H.~Lin, S.~J. Kim, S.~S{\"u}sstrunk, and M.~S. Brown, ``Revisiting radiometric
  calibration for color computer vision,'' in {\em 2011 International
  Conference on Computer Vision}, pp.~129--136, 2011.

\bibitem{RN63}
S.~J. Kim, H.~T. Lin, Z.~Lu, S.~S{\"u}sstrunk, S.~Lin, and M.~S. Brown, ``A new
  in-camera imaging model for color computer vision and its application,'' {\em
  IEEE Transactions on Pattern Analysis and Machine Intelligence}, vol.~34,
  no.~12, pp.~2289--2302, 2012.

\bibitem{RN59}
W.~Zhou, S.~Gao, L.~Zhang, and X.~Lou, ``Histogram of oriented gradients
  feature extraction from raw bayer pattern images,'' {\em IEEE Transactions on
  Circuits and Systems II: Express Briefs}, vol.~67, no.~5, pp.~946--950, 2020.

\bibitem{RN23}
M.~Buckler, S.~Jayasuriya, and A.~Sampson, ``Reconfiguring the imaging pipeline
  for computer vision,'' in {\em Proc. IEEE International Conference on
  Computer Vision}, pp.~975--984, 2017.

\bibitem{RN56}
H.~Blasinski, J.~Farrell, T.~Lian, Z.~Liu, and B.~Wandell, ``Optimizing image
  acquisition systems for autonomous driving,'' {\em Electronic Imaging},
  vol.~2018, pp.~1--7, Jan. 2018.

\bibitem{RN57}
Z.~Y. Liu, L.~Trisha, F.~Joyce, and W.~Brian, ``Neural network generalization:
  The impact of camera parameters,'' {\em arXiv}, 2019.

\bibitem{RN55}
A.~Omid-Zohoor, C.~Young, D.~Ta, and B.~Murmann, ``Toward always-on mobile
  object detection: Energy versus performance tradeoffs for embedded hog
  feature extraction,'' {\em IEEE Transactions on Circuits and Systems for
  Video Technology}, vol.~28, pp.~1102--1115, May 2018.

\bibitem{RN64}
W.~Zhou, L.~Zhang, S.~Gao, and X.~Lou, ``Gradient-based feature extraction from
  raw bayer pattern images,'' {\em IEEE Transactions on Image Processing},
  vol.~30, pp.~5122--5137, 2021.

\bibitem{RN65}
Google, ``{Camera2}.'' [Online].
  Available:\url{https://developer.android.com/training/camera2}.
\newblock Accessed May 29, 2021.

\bibitem{RN75}
J.~Deng, W.~Dong, R.~Socher, L.~Li, K.~Li, and L.~FeiFei, ``Imagenet: A
  large-scale hierarchical image database,'' in {\em Proc. IEEE Conference on
  Computer Vision and Pattern Recognition}, pp.~248--255, IEEE, 2009.

\bibitem{RN78}
C.~Yang, ``In-sensor computing for machine vision,'' {\em Nature}, vol.~579,
  no.~7797, pp.~32--33, 2020.

\bibitem{RN70}
E.~S. Lubana, R.~P. Dick, V.~Aggarwal, and P.~M. Pradhan, ``Minimalistic image
  signal processing for deep learning applications,'' in {\em Proc. IEEE
  International Conference on Image Processing}, pp.~4165--4169, 2019.

\bibitem{RN66}
P.~Isola, J.-Y. Zhu, T.~Zhou, and A.~A. Efros, ``Image-to-image translation
  with conditional adversarial networks,'' in {\em Proc. IEEE Conference on
  Computer Vision and Pattern Recognition}, pp.~1125--1134, 2017.

\bibitem{RN67}
J.-Y. Zhu, T.~Park, P.~Isola, and A.~A. Efros, ``Unpaired image-to-image
  translation using cycle-consistent adversarial networks,'' in {\em Proc. IEEE
  International Conference on Computer Vision}, pp.~2223--2232, 2017.

\bibitem{RN68}
T.~Kim, M.~Cha, H.~Kim, J.~K. Lee, and J.~Kim, ``Learning to discover
  cross-domain relations with generative adversarial networks,'' in {\em Proc.
  International Conference on Machine Learning}, pp.~1857--1865, 2017.

\bibitem{RN69}
T.~Wang, M.~Liu, J.-Y. Zhu, A.~Tao, J.~Kautz, and B.~Catanzaro,
  ``High-resolution image synthesis and semantic manipulation with conditional
  gans,'' in {\em Proc. IEEE conference on computer vision and pattern
  recognition}, pp.~8798--8807, 2018.

\bibitem{RN71}
P.~E. Debevec and J.~Malik, ``Recovering high dynamic range radiance maps from
  photographs,'' in {\em Proc. ACM Special Interest Group for Computer
  GRAPHICS}, pp.~1--10, 2008.

\bibitem{RN72}
S.~Mann and R.~W. Picard, ``On being `undigital' with digital cameras:
  Extending dynamic range by combining differently exposed pictures,'' in {\em
  Proc. Imaging Science and Technology}, pp.~442--448, 1995.

\bibitem{RN73}
T.~Mitsunaga and S.~Nayar, ``Radiometric self calibration,'' in {\em Proc. IEEE
  Computer Society Conference on Computer Vision and Pattern Recognition},
  vol.~1, pp.~374--380, 1999.

\bibitem{RN74}
D.~R. Bull, ``Digital picture formats and representations,'' in {\em
  Communicating Pictures}, pp.~99--132, Oxford: Academic Press, 2014.

\bibitem{RN76}
L.~Deng, ``The mnist database of handwritten digit images for machine learning
  research,'' {\em IEEE Signal Processing Magazine}, vol.~29, no.~6,
  pp.~141--142, 2012.

\bibitem{RN77}
A.~Krizhevsky, G.~Hinton, {\em et~al.}, ``Learning multiple layers of features
  from tiny images,'' 2009.

\bibitem{RN80}
M.~Mirza and S.~Osindero, ``Conditional generative adversarial nets,'' {\em
  arXiv preprint arXiv:1411.1784}, 2014.

\bibitem{RN93}
N.~Tran, V.~Tran, N.~Nguyen, T.~Nguyen, and N.~Cheung, ``On data augmentation
  for {GAN} training,'' {\em IEEE Transactions on Image Processing}, vol.~30,
  pp.~1882--1897, 2021.

\bibitem{RN1}
O.~Losson, L.~Macaire, and Y.~Yang, ``Comparison of color demosaicing
  methods,'' {\em Advances in Imaging and Electron Physics}, vol.~162,
  pp.~173--265, 2010.

\bibitem{RN81}
A.~Krizhevsky, I.~Sutskever, and G.~Hinton, ``Imagenet classification with deep
  convolutional neural networks,'' {\em Proc. Advances in Neural Information
  Processing Systems}, vol.~25, no.~2, 2012.

\bibitem{RN82}
K.~Simonyan and A.~Zisserman, ``Very deep convolutional networks for
  large-scale image recognition,'' {\em arXiv preprint arXiv:1409.1556}, 2014.

\bibitem{RN83}
C.~Szegedy, W.~Liu, Y.~Jia, P.~Sermanet, S.~Reed, D.~Anguelov, D.~Erhan,
  V.~Vanhoucke, and A.~Rabinovich, ``Going deeper with convolutions,'' in {\em
  Proc. IEEE Conference on Computer Vision and Pattern Recognition}, pp.~1--9,
  2015.

\bibitem{RN84}
K.~He, X.~Zhang, S.~Ren, and J.~Sun, ``Deep residual learning for image
  recognition,'' in {\em Proc. IEEE Conference on Computer Vision and Pattern
  Recognition}, pp.~770--778, 2016.

\bibitem{RN85}
S.~Ren, K.~He, R.~Girshick, and J.~Sun, ``Faster {R-CNN}: towards real-time
  object detection with region proposal networks,'' {\em IEEE transactions on
  pattern analysis and machine intelligence}, vol.~39, no.~6, pp.~1137--1149,
  2016.

\bibitem{RN86}
W.~Liu, D.~Anguelov, D.~Erhan, C.~Szegedy, S.~Reed, C.-Y. Fu, and A.~C. Berg,
  ``{SSD}: Single shot multibox detector,'' in {\em European Conference on
  Computer Vision}, pp.~21--37, 2016.

\bibitem{RN90}
T.~Ross and G.~Doll{\'a}r, ``Focal loss for dense object detection,'' in {\em
  Proc. IEEE Conference on Computer Vision and Pattern Recognition},
  pp.~2980--2988, IEEE, 2017.

\bibitem{RN89}
J.~Redmon and A.~Farhadi, ``Yolov3: An incremental improvement,'' {\em arXiv
  preprint arXiv:1804.02767}, 2018.

\bibitem{RN91}
Y.~Zhao, L.~Po, T.~Zhang, Z.~Liao, {\em et~al.}, ``Saliency map-aided
  generative adversarial network for raw to rgb mapping,'' in {\em Proc.
  IEEE/CVF International Conference on Computer Vision Workshop},
  pp.~3449--3457, 2019.

\bibitem{RN94}
T.~Karras, S.~Laine, M.~Aittala, J.~Hellsten, J.~Lehtinen, and T.~Aila,
  ``Analyzing and improving the image quality of stylegan,'' in {\em Proc. the
  IEEE/CVF Conference on Computer Vision and Pattern Recognition},
  pp.~8110--8119, 2020.

\bibitem{RN95}
A.~Odena, V.~Dumoulin, and C.~Olah, ``Deconvolution and checkerboard
  artifacts,'' {\em Distill}, vol.~1, no.~10, p.~e3, 2016.

\bibitem{RN96}
W.~Shi, J.~Caballero, F.~Husz{\'a}r, J.~Totz, A.~P. Aitken, R.~Bishop,
  D.~Rueckert, and Z.~Wang, ``Real-time single image and video super-resolution
  using an efficient sub-pixel convolutional neural network,'' in {\em Proc.
  IEEE Conference on Computer Vision and Pattern Recognition}, pp.~1874--1883,
  2016.

\bibitem{RN109}
K.~Simonyan and A.~Zisserman, ``Very deep convolutional networks for
  large-scale image recognition,'' {\em arXiv preprint arXiv:1409.1556}, 2014.

\bibitem{RN100}
A.~Ignatov, R.~Timofte, S.-J. Ko, {\em et~al.}, ``{AIM} 2019 challenge on raw
  to rgb mapping: Methods and results,'' in {\em Proc. IEEE/CVF International
  Conference on Computer Vision Workshop}, pp.~3584--3590, 2019.

\bibitem{RN99}
M.~Everingham, L.~Van~Gool, C.~K.~I. Williams, J.~Winn, and A.~Zisserman, ``The
  {PASCAL} {V}isual {O}bject {C}lasses {C}hallenge 2012 {R}esults.''
  http://www.pascal-network.org/challenges/VOC/voc2012/workshop/index.html.

\bibitem{RN105}
M.~Heusel, H.~Ramsauer, T.~Unterthiner, B.~Nessler, and S.~Hochreiter, ``{GANs}
  trained by a two time-scale update rule converge to a local nash
  equilibrium,'' in {\em Proc. Advances in Neural Information Processing
  Systems}, vol.~30, 2017.

\bibitem{RN106}
M.~Seitzer, ``{pytorch-fid: FID Score for PyTorch}.''
  \url{https://github.com/mseitzer/pytorch-fid}, August 2020.
\newblock Version 0.1.1.

\bibitem{RN16}
Z.~Wang, A.~C. Bovik, H.~R. Sheikh, E.~P. Simoncelli, {\em et~al.}, ``Image
  quality assessment: from error visibility to structural similarity,'' {\em
  IEEE Transactions on Image Processing}, vol.~13, no.~4, pp.~600--612, 2004.

\bibitem{RN108}
L.~Chang and Y.-P. Tan, ``Hybrid color filter array demosaicking for effective
  artifact suppression,'' {\em J. Electronic Imaging}, vol.~15, p.~013003,
  2006.

\bibitem{RN107}
X.~Xiao-Zhao, C.~Yi-Heng, L.~Xiao-Min, L.~Chang-Jiang, and C.~Lan-Sun,
  ``Improved grey world color correction algorithms,'' {\em Acta Photonica
  Sinica}, vol.~39, no.~3, p.~559, 2010.

\end{thebibliography}

\end{document}